\documentclass[aps,pra,showpacs,floatfix,twocolumn]{revtex4}
\usepackage{graphicx}
\usepackage{dcolumn}

\begin{document}

\title {Relativistic many-body calculations of excitation
energies and transition rates in ytterbium-like ions}

\author{U. I. Safronova}
\email{usafrono@nd.edu}
\author{W. R. Johnson}
\email{johnson@nd.edu}
\homepage{www.nd.edu/~johnson}
\author{M. S. Safronova}
\email{msafrono@nd.edu}
\altaffiliation[Current address:]
{Electron and Optical Physics
Division, National Institute of Standards and Technology,
Gaithersburg, MD, 20899-8410}
\affiliation{
Department of Physics, 225 Nieuwland Science Hall\\
University of Notre Dame, Notre Dame, IN 46566}

\author{J. R. Albritton}
\affiliation{Lawrence Livermore National Laboratory, PO Box 808,
Livermore, CA 94551}

\date{\today}

\begin{abstract}

Excitation energies, oscillator strengths, and transition
rates are calculated for
$(5d^2+5d6s+6s^2)$--$(5d6p+5d5f+6s6p)$ electric dipole transitions
in Yb-like ions with nuclear charges $Z$ ranging from 72 to 100.
Relativistic many-body perturbation theory (RMBPT), including the
retarded Breit interaction, is used to evaluate retarded E1 matrix elements
in length and velocity forms. The calculations start from a
[Xe]$4f^{14}$ core Dirac-Fock potential. First-order RMBPT is used to
obtain intermediate coupling coefficients, and second-order
RMBPT is used to determine matrix elements. A detailed
discussion of the various contributions to energy levels and
dipole matrix elements is given for ytterbium like rhenium,
$Z$=75. The resulting transition energies  are compared with
experimental values and with results from other recent calculations.
Trends of excitation energies, line strengths,  oscillator
strengths, and transition rates as functions of nuclear charge $Z$
are shown graphically for selected states and transitions.
 These calculations are presented as a
theoretical benchmark for comparison with experiment and theory.

\end{abstract}



\maketitle

\section{Introduction}

We report results of {\it ab initio} calculations of
excitation energies, oscillator strengths, and transition
rates in Yb-like ions with nuclear charges $Z$ ranging from
72 to 100. The ions considered here, starting from doubly ionized Hf~III,
all have $5d^2$ ground states. We  do not consider Yb~I and Lu~II 
both of which have a $6s^2$ ground state configuration.
In recent publications \cite{sugar,hof,kila,kilb,gayasov,churilov,optk}, the
spectra of Re~VI, Os~VII, and Ir~VIII were studied and energies levels of
the $5d^2$, $5d6s$, $5d6p$, $5d5f$ and $6s6p$
configurations were determined. The Cowan Hartree-Fock code \cite{cowan} 
with relativistic and correlation options was used in 
Refs.~\cite{sugar,hof,kila,kilb,gayasov,churilov,optk}
to calculate energy levels and to carry out least-squares adjustments of
energies.

Although we do not consider Yb~I and Lu~II here, it should be noted
that \citet{porsev} recently carried out elaborate calculations of
electric-dipole amplitudes in atomic ytterbium. 
Moreover, \citet{nist} listed
energies for 249 levels of Yb~I and 40 levels of Lu~II.

In the present paper, we use relativistic  many-body perturbation
theory (RMBPT) to determine energies of the 14 even-parity $5d^2$,
$5d6s$, and $6s^2$ states and
the 36 odd-parity  $5d6p$, $5d5f$, and $6s6p$
states for Yb-like ions.  We illustrate our calculation with a detailed
study of Re~VI, $Z$= 75. Our first-order RMBPT calculations 
include both the Coulomb and retarded Breit interactions but our
second-order calculations are limited to the Coulomb interaction only.

Reduced matrix elements, line strengths, oscillator strengths, and
transition rates are determined for all allowed and forbidden 
electric-dipole transitions between even-parity ($5d^2+5d6s+6s^2$) 
and odd-parity ($5d6p+5d5f+6s6p$) states.
Retarded E1 matrix elements are evaluated in both length and
velocity forms. RMBPT calculations that start from a local
potential are gauge independent order-by-order,
provided ``derivative terms'' are included
in second- and higher-order matrix elements and careful
attention is paid to negative-energy states. The present
calculations start from a nonlocal
[Xe]$4f^{14}$ Dirac-Fock (DF) potential and consequently give gauge-dependent
transition matrix elements. Second-order correlation corrections
compensate almost exactly for the gauge dependence of the
first-order matrix elements and lead to corrected matrix elements
that differ by less than 5\% in length and velocity forms
for all of the ions considered here.

Energies from the present calculation agree well with results given in
Refs.~\cite{sugar,hof,kila,kilb,gayasov,churilov,optk} for low-lying levels,
but disagree substantially for various highly-excited levels,
as discussed later.

\begin{table}
\caption{\label{tab0} Possible two-particle states in the Yb-like
ions}
\begin{ruledtabular}
\begin{tabular}{llll}
\multicolumn{1}{c}{$jj$ coupling}& \multicolumn{1}{c}{$LS$
coupling}& \multicolumn{1}{c}{$jj$ coupling}&
\multicolumn{1}{c}{$LS$ coupling}\\
\hline
$5d_{3/2}5d_{3/2}(0)$&$ 5d^2\ ^3P_0$&$5d_{3/2}6p_{1/2}(2)$&$ 5d6p\ ^3F_2$\\
$5d_{5/2}5d_{5/2}(0)$&$ 5d^2\ ^1S_0$&$5d_{3/2}6p_{3/2}(2)$&$ 5d6p\ ^3D_2$\\
$6s_{1/2}6s_{1/2}(0)$&$ 6s^2\ ^1S_0$&$5d_{5/2}6p_{1/2}(2)$&$ 5d6p\ ^1D_2$\\
                     &              &$5d_{5/2}6p_{3/2}(2)$&$ 5d6p\ ^3P_2$\\
$5d_{3/2}5d_{5/2}(1)$&$ 5d^2\ ^3P_1$&$5d_{3/2}5f_{5/2}(2)$&$ 6s6p\ ^3P_2$\\
$5d_{3/2}6s_{1/2}(1)$&$ 5d6s\ ^3D_1$&$5d_{3/2}5f_{7/2}(2)$&$ 5d5f\ ^3F_2$\\
                     &              &$5d_{5/2}5f_{5/2}(2)$&$ 5d5f\ ^1D_2$\\
$5d_{3/2}5d_{3/2}(2)$&$ 5d^2\ ^3F_2$&$5d_{5/2}5f_{7/2}(2)$&$ 5d5f\ ^3D_2$\\
$5d_{3/2}5d_{5/2}(2)$&$ 5d^2\ ^1D_2$&$6s_{1/2}6p_{3/2}(2)$&$ 5d5f\ ^3P_2$\\
$5d_{5/2}5d_{5/2}(2)$&$ 5d^2\ ^3P_2$&                     &            \\
$5d_{3/2}6s_{1/2}(2)$&$ 5d6s\ ^3D_2$&$5d_{3/2}6p_{3/2}(3)$&$ 5d6p\ ^3F_3$\\
$5d_{5/2}6s_{1/2}(2)$&$ 5d6s\ ^1D_2$&$5d_{5/2}6p_{1/2}(3)$&$ 5d6p\ ^3D_3$\\
                     &              &$5d_{5/2}6p_{3/2}(3)$&$ 5d6p\ ^1F_3$\\
$5d_{3/2}5d_{5/2}(3)$&$ 5d^2\ ^3F_3$&$5d_{3/2}5f_{5/2}(3)$&$ 5d5f\ ^3F_3$\\
$5d_{5/2}6s_{1/2}(3)$&$ 5d6s\ ^3D_3$&$5d_{3/2}5f_{7/2}(3)$&$ 5d5f\ ^3G_3$\\
                     &              &$5d_{5/2}5f_{5/2}(3)$&$ 5d5f\ ^3D_3$\\
$5d_{3/2}5d_{5/2}(4)$&$ 5d^2\ ^3F_4$&$5d_{5/2}5f_{7/2}(3)$&$ 5d5f\ ^1F_3$\\
$5d_{5/2}5d_{5/2}(4)$&$ 5d^2\ ^1G_4$&                     &             \\
                     &              &$5d_{5/2}6p_{3/2}(4)$&$ 5d6p\ ^3F_4$\\
$5d_{3/2}6p_{3/2}(0)$&$ 5d6p\ ^3P_0$&$5d_{3/2}5f_{5/2}(4)$&$ 5d5f\ ^1G_4$\\
$5d_{5/2}5f_{5/2}(0)$&$ 6s6p\ ^3P_0$&$5d_{3/2}5f_{7/2}(4)$&$ 5d5f\ ^3H_4$\\
$6s_{1/2}6p_{1/2}(0)$&$ 5d5f\ ^3P_0$&$5d_{5/2}5f_{5/2}(4)$&$ 5d5f\ ^3F_4$\\
                     &              &$5d_{5/2}5f_{7/2}(4)$&$ 5d5f\ ^3G_4$\\
$5d_{3/2}6p_{1/2}(1)$&$ 5d6p\ ^3D_1$&                     &             \\
$5d_{3/2}6p_{3/2}(1)$&$ 5d6p\ ^3P_1$&$5d_{3/2}5f_{7/2}(5)$&$ 5d5f\ ^3H_5$\\
$5d_{5/2}6p_{3/2}(1)$&$ 5d6p\ ^1P_1$&$5d_{5/2}5f_{5/2}(5)$&$ 5d5f\ ^3G_5$\\
$5d_{3/2}5f_{5/2}(1)$&$ 6s6p\ ^3P_1$&$5d_{5/2}5f_{7/2}(5)$&$ 5d5f\ ^1H_5$\\
$5d_{5/2}5f_{5/2}(1)$&$ 5d5f\ ^3D_1$&                     &            \\
$5d_{5/2}5f_{7/2}(1)$&$ 6s6p\ ^1P_1$&$5d_{5/2}5f_{7/2}(6)$&$ 5d5f\ ^3H_6$\\
$6s_{1/2}6p_{1/2}(1)$&$ 5d5f\ ^3P_1$&                     &             \\
$6s_{1/2}6p_{3/2}(1)$&$ 5d5f\ ^1P_1$&                     &             \\
\end{tabular}
\end{ruledtabular}
\end{table}

\begin{table}
\caption{\label{tab1} Contributions to energy matrices $E[Q,Q']$
(a.u.)\ for odd-parity states $Q$=$n_1l_1j_1n_2l_2j_2(J)$, $Q'$=
$n_3l_3j_3n_4l_4j_4(J)$ with $J$=0  before diagonalization in  the
case of Yb-like rhenium, $Z=75$.}
\begin{ruledtabular}
\begin{tabular}{lrrrr}
\multicolumn{1}{c}{$Q,\ Q'$} & \multicolumn{1}{c}{$E^{(0)}$} &
\multicolumn{1}{c}{$E^{(1)}$} & \multicolumn{1}{c}{$B^{(1)}$}&
\multicolumn{1}{c}{$E^{(2)}$} \\
\hline
$5d_{3/2}6p_{3/2},\ 5d_{3/2}6p_{3/2}$& -3.99398 & 0.42731 & 0.00679 & -0.14018 \\
$5d_{5/2}5f_{5/2},\ 5d_{5/2}5f_{5/2}$& -3.42978 & 0.44889 & 0.00430 & -0.16664 \\
$6s_{1/2}6p_{1/2},\ 6s_{1/2}6p_{1/2}$& -3.64181 & 0.29737 & 0.00591 & -0.10004 \\
$5d_{3/2}6p_{3/2},\ 5d_{5/2}5f_{5/2}$&          &-0.01871 & 0.00003 &  0.01465 \\
$5d_{5/2}5f_{5/2},\ 5d_{3/2}6p_{3/2}$&          &-0.01871 & 0.00003 &  0.00965 \\
$5d_{3/2}6p_{3/2},\ 6s_{1/2}6p_{1/2}$&          & 0.02191 &-0.00001 & -0.01278 \\
$6s_{1/2}6p_{1/2},\ 5d_{3/2}6p_{3/2}$&          & 0.02191 &-0.00001 & -0.00996 \\
$5d_{5/2}5f_{5/2},\ 6s_{1/2}6p_{1/2}$&          &-0.01992 &-0.00001 &  0.00221 \\
$6s_{1/2}6p_{1/2},\ 5d_{5/2}5f_{5/2}$&          &-0.01992 &-0.00001 &  0.00252 \\
\end{tabular}
\end{ruledtabular}
\end{table}

\section {Method}
The RMBPT formalism developed previously
\cite{be2e,neutr,be2t,mg,d2} for Be-, Mg-, and Ca-like ions is used
here to describe perturbed wave functions, to obtain the
second-order energies \cite{be2e}, and to evaluate first- and
second-order transition matrix elements \cite{be2t}.  
Ions of the Yb sequence, starting from the Hf III ion \cite{klin} 
and continuing onward have a $5d^2$ ground state. 
This is similar to the previously
studied Ca sequence \cite{d2}, where ions starting from Ti III 
have a $3d^2$ ground state.
The primary differences between
calculations for Ca-like and Yb-like ions arise from the increased
number of the orbitals in the DF core potential,
[Xe]$4f^{14}$ instead of [Ar]  
([Ar] = $1s^22s^22p^63s^23p^6$
and [Xe] = [Ar]\,$3d^{10}4s^24p^64d^{10}5s^25p^6$), and the strong mixing between both
even-parity ($5d^2+5d6s+6s^2$) and odd-parity ($5d6p+5d5f+6s6p$) states. 
These differences lead to much
more laborious numerical calculations. The calculations are
carried out using  sets of DF basis 
orbitals that are linear combinations of B-splines. 
These B-spline basis
orbitals are determined using the method described in
Ref.~\cite{2wrj}. We use 40 B-splines of order 8 for each
single-particle angular momentum state and we include all orbitals
with orbital angular momentum $l \leq 7$ in our basis set.

\subsection{Model space}

The model spaces for the ($5d^2+5d6s+6s^2$) and ($5d6p+5d5f+6s6p$)
complexes in Yb-like ions have 14 even-parity states and 36
odd-parity states, respectively. 
These states are summarized in Table~\ref{tab0},
where both $jj$ and $LS$ designations are given. When starting
calculations from DF wave functions, it is
natural to use $jj$ designations for uncoupled 
matrix elements; however, neither $jj$- nor $LS$-coupling
describes {\em physical} states properly, except for the
single-configuration state $5d_{5/2}5f_{7/2}(6) \equiv 5d5f\ ^3\!
H_6$.
The strong mixing between $5d6p$, $5d5f$, and $6s6p$ states was
discussed previously in
Refs.~\cite{sugar,hof,kila,kilb,gayasov,churilov,optk}.

\begin{table}
\caption{\label{taben} Energy of $5d^2$, $5d6s$, $6s^2$, $5d6p$,
$5d5f$, and $6s6p$
 states in Yb-like Re$^{+5}$ (cm$^{-1}$).
 Notation: $E^{(0+1)}$=$E^{(0)}$+$E^{(1)}$+$B^{(1)}$.}
\begin{ruledtabular}
\begin{tabular}{lrrrrrr}
\multicolumn{1}{c}{ }& \multicolumn{1}{c}{$E^{(0+1)}$}&
\multicolumn{1}{c}{$E^{(2)}$}& \multicolumn{1}{c}{$E^{(\rm
{tot})}$}& \multicolumn{1}{c}{$E^{(0+1)}$}&
\multicolumn{1}{c}{$E^{(2)}$}&
\multicolumn{1}{c}{$E^{(\rm {exc})}$}\\
\multicolumn{1}{c}{Level}& \multicolumn{3}{c}{Absolute energies}&
\multicolumn{3}{c}{Excitation energies}\\
\hline
$5d^2\ ^3F_2$&-1184024&   -40817& -1224841&  0    &  0   &   0   \\
$5d^2\ ^3F_3$&-1177178&   -39172& -1216350&   6847&  1645&   8491\\
$5d^2\ ^3F_4$&-1170966&   -38780& -1209746&  13058&  2037&  15095\\
$5d^2\ ^3P_0$&-1167037&   -43309& -1210346&  16987& -2493&  14495\\
$5d^2\ ^1D_2$&-1166402&   -41572& -1207975&  17622&  -756&  16866\\
$5d^2\ ^3P_1$&-1163473&   -41812& -1205285&  20551&  -995&  19556\\
$5d^2\ ^3P_2$&-1156318&   -40152& -1196470&  27706&   665&  28371\\
$5d^2\ ^1G_4$&-1155966&   -42068& -1198034&  28059& -1251&  26807\\
$5d^2\ ^1S_0$&-1127568&   -46217& -1173785&  56456& -5400&  51056\\
$5d6s\ ^3D_1$&-1097044&   -35141& -1132185&  86980&  5675&  92656\\
$5d6s\ ^3D_2$&-1094845&   -35442& -1130287&  89180&  5375&  94554\\
$5d6s\ ^3D_3$&-1087729&   -33669& -1121397&  96296&  7148& 103444\\
$5d6s\ ^1D_2$&-1079142&   -36325& -1115467& 104883&  4491& 109374\\[0.4pc]
$5d6p\ ^3F_2$&-1030357&   -30866& -1061223& 153667&  9951& 163618\\
$5d6p\ ^3D_1$&-1024847&   -33457& -1058303& 159178&  7360& 166538\\
$5d6p\ ^3D_2$&-1018155&   -30518& -1048673& 165869& 10299& 176168\\
$5d6p\ ^3F_3$&-1017295&   -30334& -1047629& 166729& 10483& 177212\\
$5d6p\ ^1D_2$&-1012677&   -29311& -1041988& 171347& 11505& 182853\\
$5d6p\ ^3D_3$&-1007838&   -30332& -1038169& 176187& 10485& 186672\\
$5d6p\ ^3P_1$&-1005870&   -31934& -1037804& 178154&  8882& 187037\\
$5d6p\ ^3P_0$&-1004380&   -31262& -1035642& 179644&  9554& 189199\\
$5d6p\ ^3F_4$&-1001386&   -28009& -1029395& 182639& 12807& 195446\\
$5d6p\ ^3P_2$& -999246&   -29109& -1028355& 184778& 11708& 196486\\
$5d6p\ ^1F_3$& -995855&   -31244& -1027099& 188169&  9573& 197742\\
$6s^2\ ^1S_0$& -991963&   -34225& -1026189& 192061&  6592& 198652\\
$5d6p\ ^3P_1$& -989455&   -34880& -1024335& 194569&  5936& 200506\\[0.4pc]
$6s6p\ ^3P_0$& -941422&   -21782&  -963204& 242603& 19035& 261637\\
$6s6p\ ^3P_1$& -936597&   -22693&  -959291& 247427& 18123& 265550\\
$6s6p\ ^3P_2$& -921614&   -21582&  -943197& 262410& 19234& 281644\\
$5d5f\ ^1G_4$& -914890&   -27340&  -942229& 269135& 13477& 282611\\
$5d5f\ ^3H_4$& -911078&   -28960&  -940038& 272946& 11857& 284803\\
$5d5f\ ^3H_5$& -910564&   -26917&  -937481& 273460& 13900& 287360\\
$5d5f\ ^3F_2$& -909842&   -32013&  -941855& 274182&  8804& 282986\\
$5d5f\ ^3F_3$& -907828&   -33638&  -941466& 276197&  7178& 283375\\
$5d5f\ ^3H_6$& -905206&   -23491&  -928698& 278818& 17325& 296143\\
$5d5f\ ^3F_4$& -903304&   -28769&  -932074& 280720& 12047& 292767\\
$5d5f\ ^1D_2$& -900678&   -34598&  -935275& 283347&  6219& 289565\\
$5d5f\ ^3D_1$& -898884&   -36856&  -935740& 285140&  3960& 289100\\
$5d5f\ ^3G_3$& -897436&   -41168&  -938604& 286588&  -352& 286237\\
$5d5f\ ^3D_2$& -893563&   -36014&  -929576& 290462&  4803& 295265\\
$6s6p\ ^1P_1$& -892813&   -36571&  -929383& 291211&  4246& 295457\\
$5d5f\ ^3G_4$& -892131&   -38708&  -930839& 291893&  2108& 294002\\
$5d5f\ ^3D_3$& -891083&   -38081&  -929165& 292941&  2735& 295676\\
$5d5f\ ^3G_5$& -889833&   -39291&  -929124& 294191&  1526& 295717\\
$5d5f\ ^3P_2$& -887212&   -36551&  -923763& 296812&  4266& 301078\\
$5d5f\ ^3P_1$& -886406&   -36865&  -923271& 297619&  3952& 301570\\
$5d5f\ ^3P_0$& -886124&   -36385&  -922510& 297900&  4431& 302331\\
$5d5f\ ^1F_3$& -884984&   -39786&  -924770& 299040&  1031& 300071\\
$5d5f\ ^1H_5$& -872058&   -49224&  -921282& 311966& -8407& 303559\\
$5d5f\ ^1P_1$& -866971&   -40448&  -907419& 317054&   369& 317422\\
\end{tabular}
\end{ruledtabular}
\end{table}

\subsection{Example: energy matrix for Re$^{+5}$ }
   Details of the theoretical method used
to evaluate second-order energies for ions with two valence
electrons 
are given in Refs.~\cite{be2e,neutr}  and will not be
repeated here. The energy calculations are illustrated in Table
\ref{tab1}, where we list contributions to the energies
 of odd-parity $J=0$ states of Re$^{+5}$.
We present zeroth-, first-, and second-order Coulomb
energies  $E^{(0)}$,  $E^{(1)}$, and $E^{(2)}$
together with the first-order retarded Breit
corrections $B^{(1)}$ \cite{chen}. It should be mentioned that
the difference
between first-order Breit corrections calculated with
and without retardation is less than 2\%.  As one can see
from Table~\ref{tab1}, the ratio of off-diagonal to diagonal
matrix elements is larger for second-order contributions than
for first-order contributions. Another difference between 
first- and second-order contributions concerns symmetry properties:
first-order off-diagonal matrix elements are symmetric, whereas
second-order off-diagonal matrix elements are unsymmetric
(\citet[chap.\ 9]{LM}).
Indeed, 
$E^{(2)}[Q,Q']$ and $E^{(2)}[Q',Q]$  differ in some
cases by 20-50\%  and occasionally even have opposite signs. 
The ratio of off-diagonal to diagonal matrix
elements for Breit corrections $B^{(1)}$ is much smaller
than for Coulomb corrections.

After evaluating the energy matrices,  we calculate eigenvalues
and eigenvectors for states with given values of $J$ and parity.
There are two possible methods  to  carry out the diagonalization:
either diagonalize the sum of  zeroth-  and  first-order  matrices,
then calculate the second-order contributions using the resulting
eigenvectors; or  diagonalize the sum of  the zeroth-,  first-
and  second-order matrices together. Following Ref.~\cite{neutr},
we choose the second method here.

The energy calculations are illustrated for Re$^{+5}$ in Table~\ref{taben}. 
Energies listed under the heading ``Absolute energies'' are 
given relative to the [Xe]$4f^{14}$ core,
while those listed under the heading ``Excitation energies'' 
are given relative to the $5d^{2}\ ^3F_2$ ground state. 
In the table, we present absolute zeroth- plus first-order
Coulomb and Breit energies $E^{(0+1)} = E^{(0)}+E^{(1)}+B^{(1)}$,  
absolute second-order Coulomb energies $E^{(2)}$,
and the sum $E^{(\rm {tot})}$. We also give the breakdown excitation energies
and total excitation energies $E^{(\rm {exc})}$. 
As can be seen from the table, the second-order contribution
is about 3\% of the absolute energy but accounts for 5\% -- 15\% of the
excitation energy. This table clearly illustrates the importance of
including second-order contributions.
As mentioned previously,  neither $jj$- nor $LS$-coupling
describes physical states properly; nevertheless, we use
$LS$ designations to label levels. 
We organize levels in the table according to
decreasing values of $E^{(0+1)}$.  After including second-order
corrections, the present ordering differs in some cases from 
the ordering according to decreasing 
$E^{(\rm {tot})}$. Thus,  the ordering of $5d^2\ ^3F_4$
and $5d^2\ ^3P_0$ levels, $5d5f\ ^3H_5$ and $5d5f\ ^3F_2$ levels,
and $5d5f\ ^3G_4$ and $6s6p\ ^1P_1$ levels are interchanged in Re$^{+5}$
after second-order corrections are added.

Problems arising when using
different model spaces in RMBPT theory were examined by
\citet{neutr}. A major difference
between Yb-like and Be-like systems, is that we could not construct as
complete a model space for a two-electron system with a
[Xe]$4f^{14}$ core
as we did for a system with a [He] core \cite{be2e}. 
To do so would require us to include all possible two-particle states
that could be constructed from unoccupied $n$=5 and
$n$=6 orbitals in our model space. In the case of Be-like ions, the
model space is much simpler, being constructed from $n$=2 orbitals only.
A second, but related,  problem is that different model spaces
lead to different results. For example, 
energy levels of  Hg~I, calculated using RMBPT
with  ($6s6p$) and ($6s^2 + 6p^2$) model spaces 
differed from those calculated with
($6s6p +6p6d$) and ($6s^2 + 6p^2+6s6d$) model spaces
by about 500 cm$^{-1}$ \cite{neutr}. 
We confirm this result here comparing calculations of $5p6d$ 
energy levels starting
from a $5p6d$ model space and those starting from 
($5p6d+ 5d5f$) or ($5p6d+ 5d5f+6s6p$) model spaces in Re$^{+5}$. 
The largest difference occurs in the 
$E^{(2)}$, which changes by 1000 cm$^{-1}$ in some cases. 
The change in $E^{(0+1)}$  is smaller by a factor of two and
has an opposite sign.
The resulting change in  $E^{(\rm {tot})}$ is about 500 cm$^{-1}$.
Similar tests of model-space dependence 
along the Yb sequence set a limit of about 500 cm$^{-1}$ 
on the accuracy of the present second-order calculations. 

\subsection{$Z$-dependence of energies}

\begin{figure*}
\centerline{\includegraphics[scale=0.3]{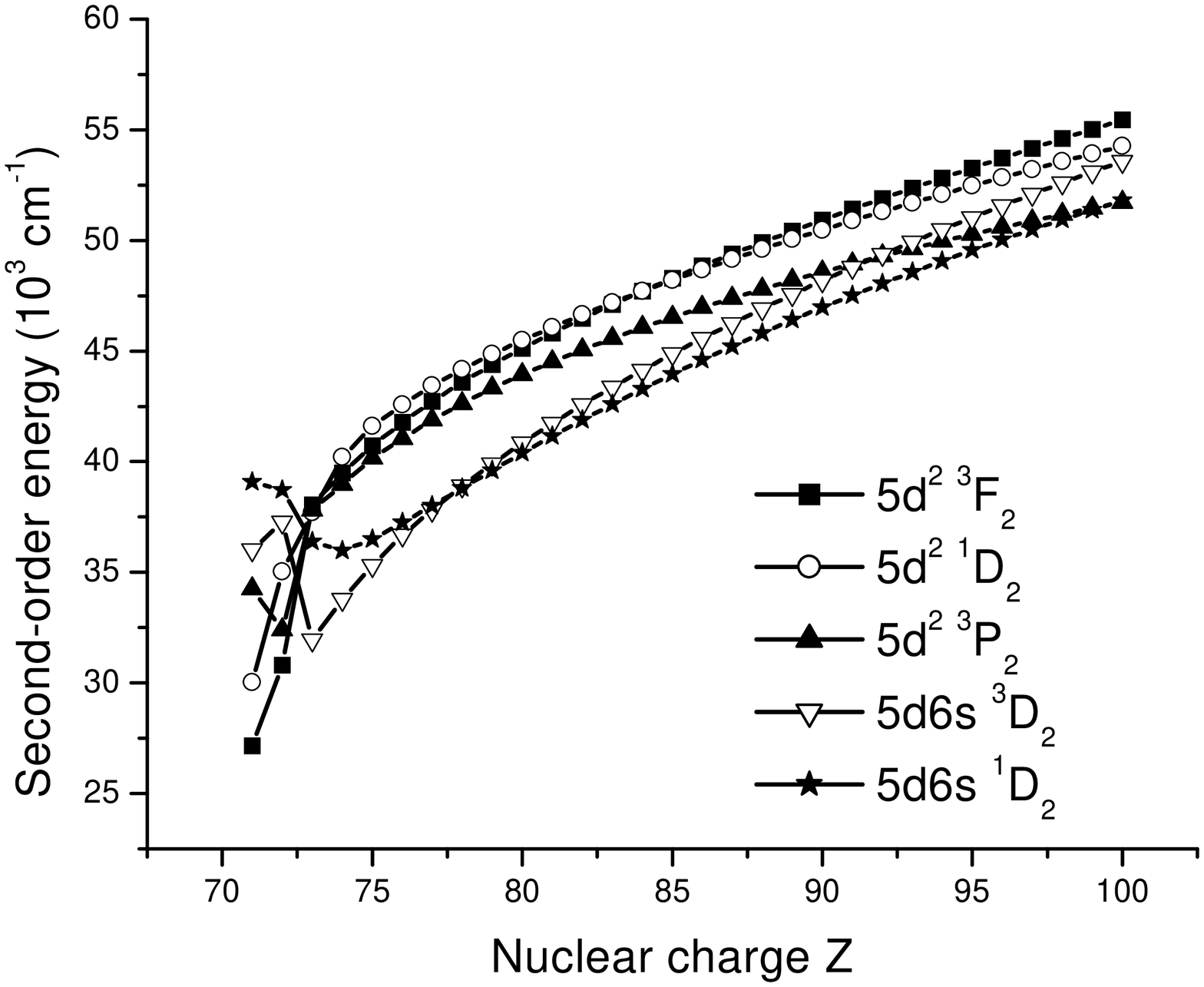}
\quad \includegraphics[scale=0.3]{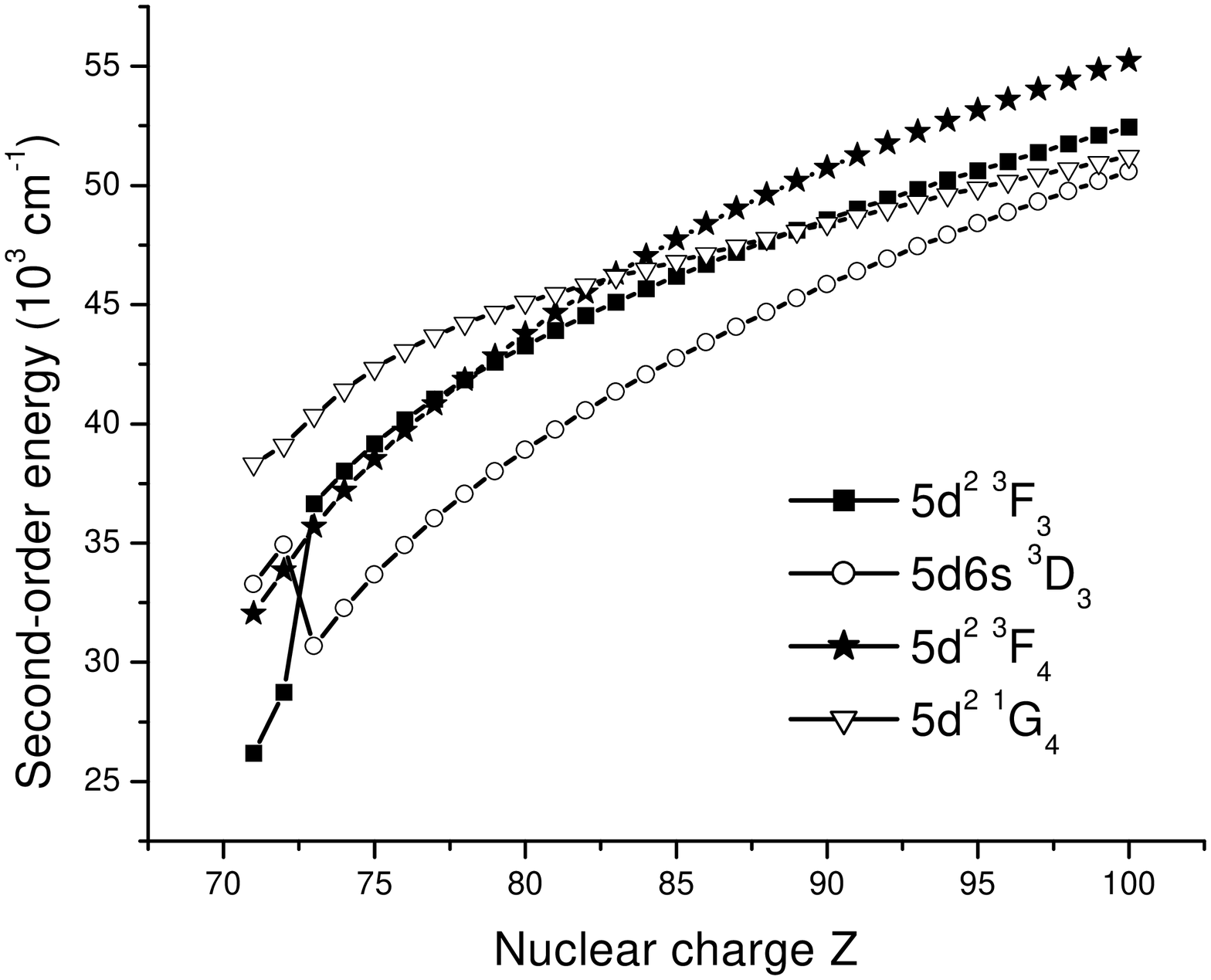}} 
\caption{$Z$-dependence of the second-order energy $E^{(2)}$ for the
  $5d^2$ and $5d6s$ energy levels}
\label{fig1}
\end{figure*}

One unique feature of the present calculations is the inclusion of 
second-order correlation corrections.
We illustrate the $Z$-dependence of the
second-order energy  $E^{(2)}$ in Fig.~\ref{fig1} 
for even-parity levels
with $J$=2 ($5d^2\ ^3F_2,\, ^1D_2,\, ^3P_2$ and $5d6s\ ^{1,3}D_2$)
and $J$=3, 4 ($5d^2\ ^3F_3,\, ^3F_4,\, ^1G_4$ and $5d6s\ ^{3}D_3$). 
As can see from this figure, the second-order energy, $E^{(2)}$
slowly increases with $Z$  in the range 3--5$\times
10^4$~cm$^{-1}$. The smooth $Z$-dependence
for these nine terms is exceptional and is not found
for other terms discussed below.

\begin{figure*}
\centerline{\includegraphics[scale=0.3]{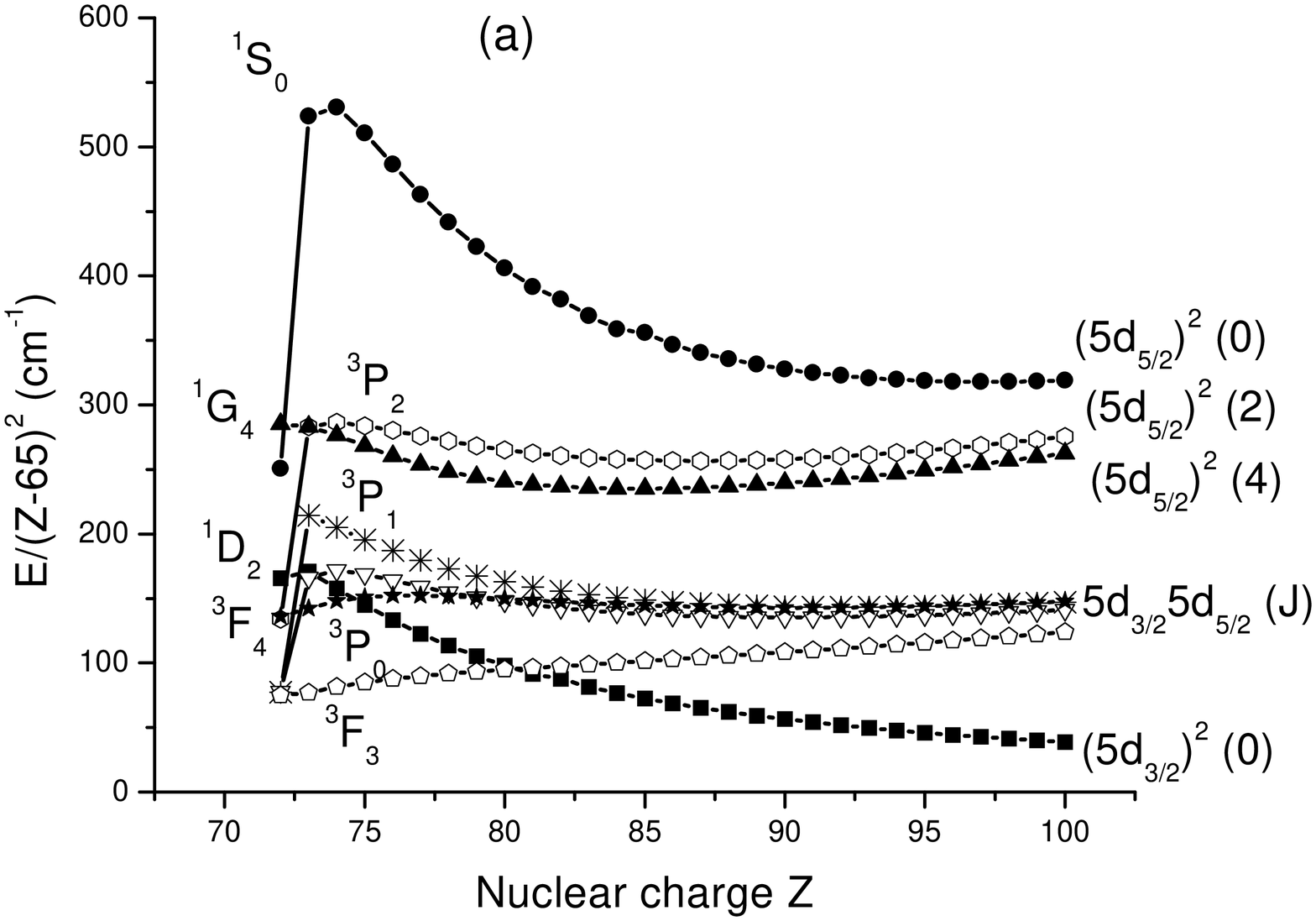}
\quad \includegraphics[scale=0.3]{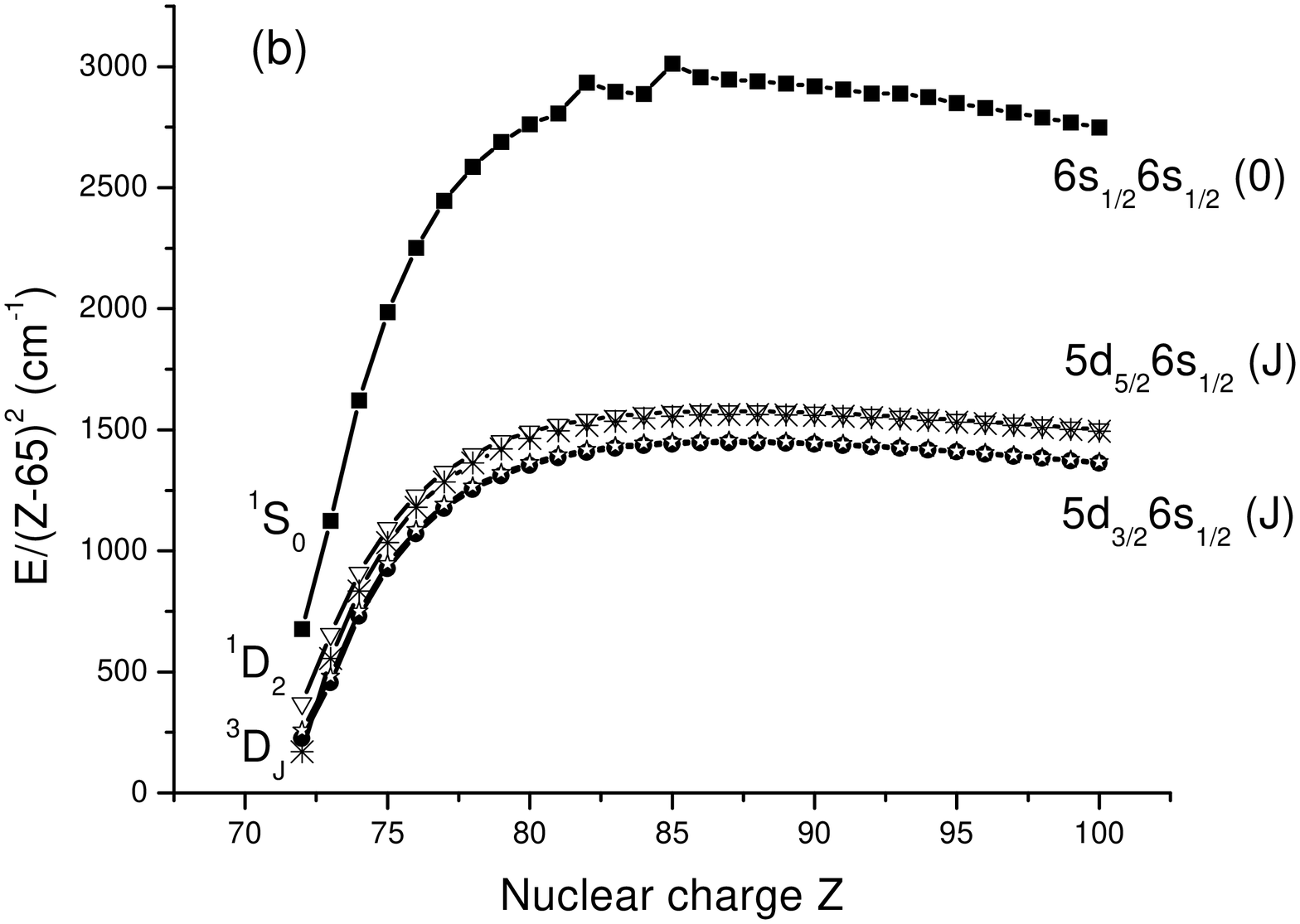}}
\caption{$Z$-dependence of the excitation energy
$E^{(\text{exc})}/(Z-65)^2$ in cm$^{-1}$.} 
for even-parity levels.
\label{en-eve}
\end{figure*}

\begin{figure*}
\centerline{\includegraphics[scale=0.3]{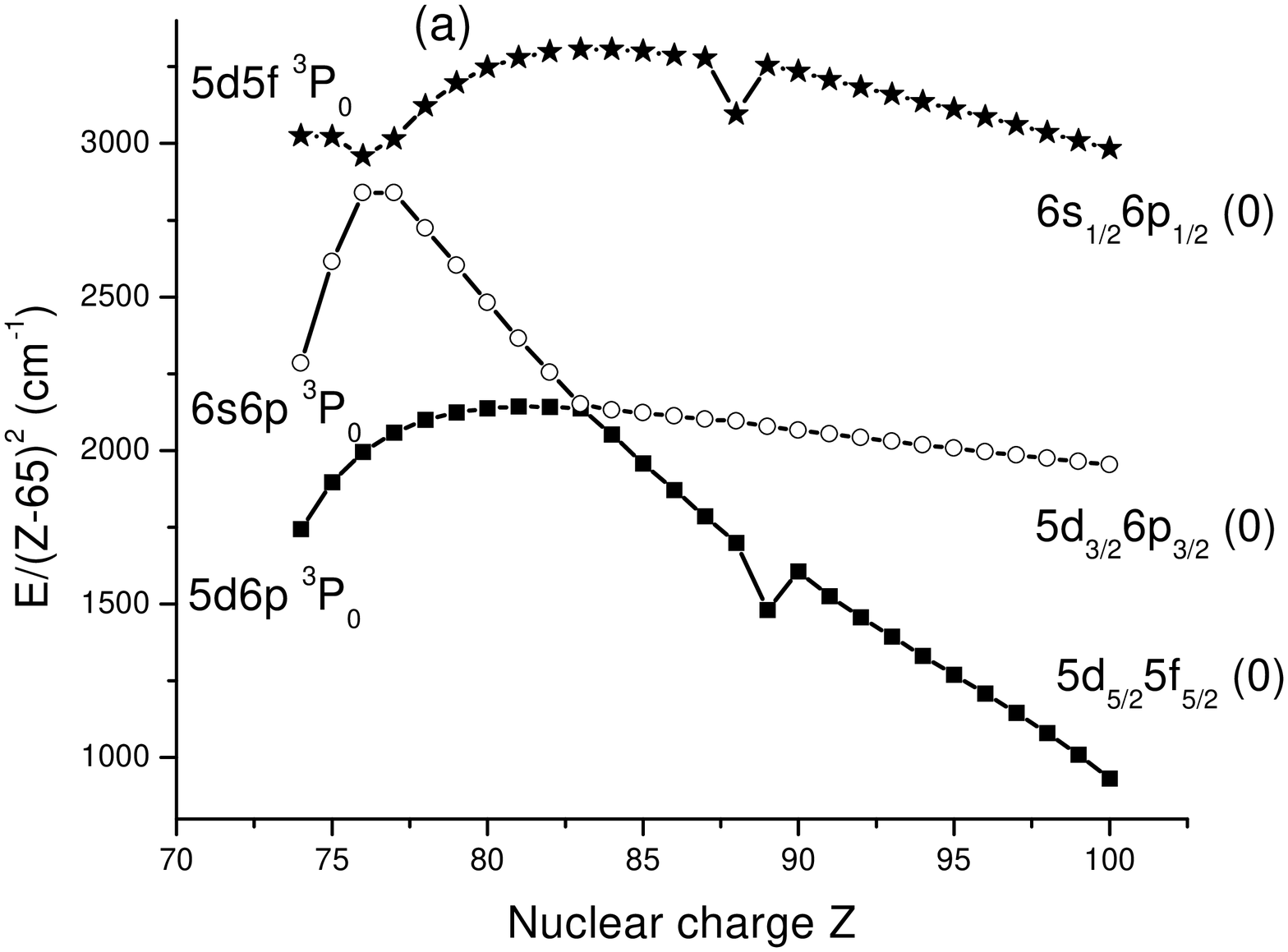}
\quad \includegraphics[scale=0.3]{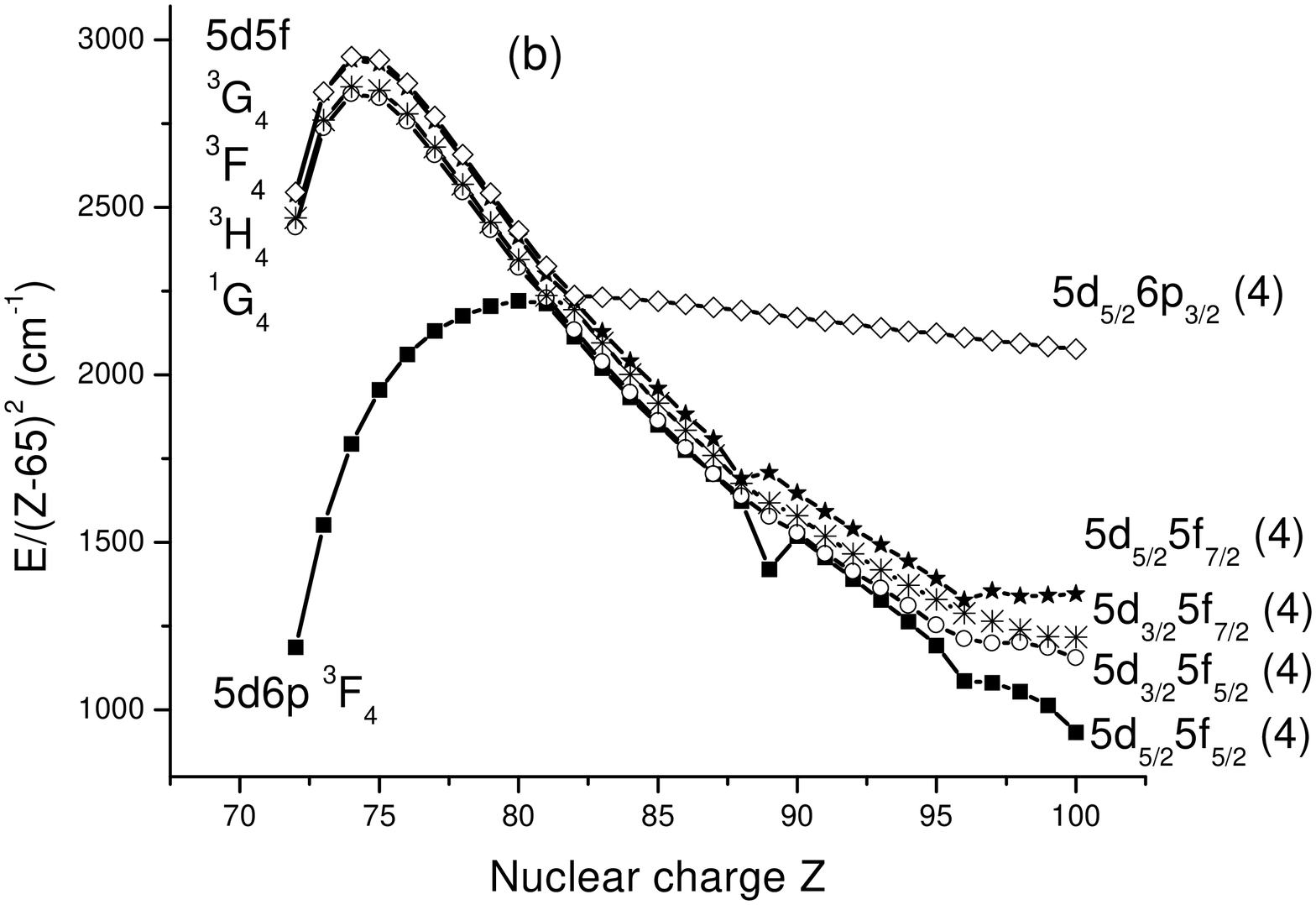}}
\caption{$Z$-dependence of the excitation energy
$E^{(\text{exc})}/(Z-65)^2$ in cm$^{-1}$.} 
for odd-parity levels.
\label{en-odd}
\end{figure*}

Excitation energies $E^{(\text{exc})}$ of even-parity and odd-parity states
relative to the $5d^2\ ^3F_2$  ground state, divided by $(Z-65)^2$, are
shown in Figs.~\ref{en-eve} and \ref{en-odd}. Both
designations are shown in these figures: $LS$ for low $Z$ and $jj$
for high $Z$.
The variation of the $5d^2$  levels with $Z$ shown in Fig.~\ref{en-eve}a. 
Strong mixing between $5d_{3/2}5d_{5/2}(J)$ and $5d_{5/2}5d_{5/2}(J)$
states with $J$=2 or 4 leads to rapid variations with $Z$ in the
corresponding Grotrian diagrams.  Note that the  $^1D_2$ and
$^3F_4$ levels cross between $Z$=79 and 80,  $^3P_0$ and $^3F_3$ levels cross 
between $Z$=80 and 81, and $^3P_1$ and $^3F_4$ levels cross between
$Z$=93 and 94.
We give excitation energies of the other five even-parity  $5d6s$  and
$6s^2$ levels in Fig.~\ref{en-eve}b. Two small sharp features in the 
$6s^2$ level occur at  $Z$=82 and $Z$=85. The origin of these 
irregularities is  discussed in Appendix A.
Energies of odd-parity
levels with $J$=0 and 4, including mixing of $5d6p$, $5d5f$ and $6s6p$ states,
are given in Fig.~\ref{en-odd}.
The sharp features in the curves
describing $5d5f$ and $6s6p$ states are similar to those 
mentioned above for even-parity
states and
discussed in Appendix A.  
Avoided level crossings for the odd-parity $J=0$ levels 
are seen Fig.~\ref{en-odd}a near $Z$=76 and $Z$=84.
Similar avoided crossings can be observed
for the odd-parity complex with $J$=4 in Fig.~\ref{en-odd}b.

\begin{table*}
\caption{Uncoupled reduced matrix elements in length $L$ and
velocity $V$ forms for even-odd parity transitions in Re$^{+5}$.}
\begin{ruledtabular}
\begin{tabular}{llrrrrrrrr}
\multicolumn{1}{c}{even-parity}& \multicolumn{1}{c}{odd-parity}&
\multicolumn{1}{c}{$Z^{(1)}_L$} &
\multicolumn{1}{c}{$Z^{(1)}_V$} & \multicolumn{1}{c}{$Z^{(2)}_L$}
& \multicolumn{1}{c}{$Z^{(2)}_V$} &
\multicolumn{1}{c}{$B^{(2)}_L$} & \multicolumn{1}{c}{$B^{(2)}_V$}
& \multicolumn{1}{c}{$P^{(\rm derv)}_L$} &
\multicolumn{1}{c}{$P^{(\rm derv)}_V$}\\[0.25pc]
\hline
$5d_{3/2}5d_{3/2}(0)$&$5d_{3/2}6p_{1/2}(1)$&  0.84736&  0.73795& -0.18176& -0.32201&  0.00105&  0.00032&  0.84733& -0.00001\\
$5d_{3/2}5d_{3/2}(0)$&$5d_{3/2}6p_{3/2}(1)$& -0.33505& -0.29206&  0.00633&  0.03976& -0.00074& -0.00053& -0.33502&  0.00003\\
$5d_{3/2}5d_{3/2}(0)$&$5d_{5/2}6p_{3/2}(1)$&  0.00000&  0.00000& -0.06416& -0.09997&  0.00002&  0.00000&  0.00000&  0.00000\\
$5d_{5/2}5d_{5/2}(0)$&$5d_{3/2}6p_{1/2}(1)$&  0.00000&  0.00000& -0.04075& -0.08423& -0.00001& -0.00002&  0.00000&  0.00000\\
$5d_{5/2}5d_{5/2}(0)$&$5d_{3/2}6p_{3/2}(1)$&  0.00000&  0.00000&  0.05057&  0.07583&  0.00000&  0.00001&  0.00000&  0.00000\\
$5d_{5/2}5d_{5/2}(0)$&$5d_{5/2}6p_{3/2}(1)$&  0.87030&  0.75077& -0.15075& -0.26297&  0.00100&  0.00055&  0.87028&  0.00003\\
$6s_{1/2}6s_{1/2}(0)$&$5d_{3/2}6p_{1/2}(1)$&  0.00000&  0.00000& -0.30143&  0.03609&  0.00006&  0.00004&  0.00000&  0.00000\\
$6s_{1/2}6s_{1/2}(0)$&$5d_{3/2}6p_{3/2}(1)$&  0.00000&  0.00000&  0.20141& -0.18719& -0.00003& -0.00001&  0.00000&  0.00000\\
$6s_{1/2}6s_{1/2}(0)$&$5d_{5/2}6p_{3/2}(1)$&  0.00000&  0.00000& -0.53534&  0.90941&  0.00006&  0.00011&  0.00000&  0.00000\\
\end{tabular}
\end{ruledtabular}
\label{tab-dip}
\end{table*}

\begin{table}
\caption{Line strengths in length $L$ and velocity $V$ forms for
even-odd-parity transitions  in Re$^{+5}$.}
\begin{ruledtabular}
\begin{tabular}{llrrrr}
\multicolumn{2}{c}{}&
\multicolumn{2}{c}{RMBPT}&
\multicolumn{2}{c}{First order}\\
\multicolumn{1}{c}{Level}&
\multicolumn{1}{c}{Level}&
\multicolumn{1}{c}{$L$} &
\multicolumn{1}{c}{$V$} &
\multicolumn{1}{c}{$L$} &
\multicolumn{1}{c}{$V$} \\[0.25pc]
\hline
 $5d^2\ ^3P_0$&$5d6p\ ^3D_1$&  0.2632&  0.2312&   0.4952&   0.3749\\
 $5d^2\ ^3P_0$&$5d6p\ ^3P_1$&  0.2424&  0.2277&   0.2673&   0.2027\\
 $5d^2\ ^3P_0$&$5d6p\ ^1P_1$&  0.0645&  0.0651&   0.0532&   0.0391\\
 $5d^2\ ^1S_0$&$5d6p\ ^3D_1$&  0.0040&  0.0032&   0.0375&   0.0288\\
 $5d^2\ ^1S_0$&$5d6p\ ^3P_1$&  0.0050&  0.0052&   0.0272&   0.0208\\
 $5d^2\ ^1S_0$&$5d6p\ ^1P_1$&  0.3261&  0.3437&   0.7024&   0.5236\\
 $5d^2\ ^3P_1$&$5d6p\ ^3P_0$&  0.5035&  0.5010&   0.5657&   0.4208\\
 $5d6s\ ^3D_1$&$5d6p\ ^3P_0$&  1.1657&  1.1792&   1.6949&   1.5356\\
 $5d^2\ ^3P_1$&$5d6p\ ^3D_1$&  0.0080&  0.0081&   0.0242&   0.0179\\
 $5d^2\ ^3P_1$&$5d6p\ ^3P_1$&  0.4143&  0.4127&   0.5065&   0.3774\\
 $5d^2\ ^3P_1$&$5d6p\ ^1P_1$&  0.0358&  0.0340&   0.0346&   0.0264\\
 $5d6s\ ^3D_1$&$5d6p\ ^3P_1$&  2.1855&  2.2762&   3.0478&   2.7590\\
 $5d^2\ ^3P_1$&$5d6p\ ^1D_2$&  0.0274&  0.0296&   0.0709&   0.0528\\
 $5d^2\ ^3P_1$&$5d6p\ ^3P_2$&  0.2803&  0.2573&   0.2959&   0.2251\\
 $5d^2\ ^3F_2$&$5d6p\ ^3P_1$&  0.1648&  0.1639&   0.1486&   0.1122\\
 $5d^2\ ^3F_2$&$5d6p\ ^3D_2$&  0.1145&  0.1145&   0.1690&   0.1262\\
 $5d6s\ ^3D_2$&$5d6p\ ^3P_2$&  1.2140&  1.2595&   1.5450&   1.3925\\
 $5d^2\ ^1D_2$&$5d6p\ ^3F_3$&  0.0380&  0.0361&   0.1735&   0.1284\\
 $5d6s\ ^3D_2$&$5d6p\ ^3F_3$&  2.4488&  2.7097&   3.3267&   3.0271\\
 $5d^2\ ^3F_3$&$5d6p\ ^1D_2$&  1.6146&  1.6111&   1.7279&   1.2837\\
 $5d^2\ ^3F_3$&$5d6p\ ^3P_2$&  0.1384&  0.1440&   0.1441&   0.1079\\
 $5d6s\ ^3D_3$&$5d6p\ ^3F_2$&  0.0020&  0.0027&   0.0026&   0.0024\\
 $5d6s\ ^3D_3$&$5d6p\ ^3P_2$&  2.7197&  2.9674&   3.8172&   3.4676\\
 $5d^2\ ^3F_3$&$5d6p\ ^3F_3$&  1.6878&  1.5330&   1.9319&   1.4599\\
 $5d^2\ ^3F_3$&$5d6p\ ^3D_3$&  0.0528&  0.0563&   0.0308&   0.0219\\
 $5d6s\ ^3D_3$&$5d6p\ ^1F_3$&  1.7322&  1.8326&   2.3084&   2.0843\\
 $5d^2\ ^3F_3$&$5d6p\ ^3F_4$&  0.1367&  0.1294&   0.1291&   0.0983\\
 $5d6s\ ^3D_3$&$5d6p\ ^3F_4$& 11.1345& 11.5018&  15.2305&  13.8004\\
 $5d^2\ ^3F_4$&$5d6p\ ^3F_3$&  0.5896&  0.5200&   0.8001&   0.6090\\
 $5d^2\ ^3F_4$&$5d6p\ ^3D_3$&  3.6963&  3.6447&   4.0769&   3.0379\\
 $5d^2\ ^3F_4$&$5d6p\ ^1F_3$&  0.1442&  0.1561&   0.1513&   0.1095\\
 $5d^2\ ^1G_4$&$5d6p\ ^3D_3$&  0.3118&  0.3027&   0.3497&   0.2616\\
 $5d^2\ ^1G_4$&$5d6p\ ^1F_3$&  5.6434&  5.5815&   6.2102&   4.6443\\
 $5d^2\ ^3F_4$&$5d6p\ ^3F_4$&  1.9537&  1.9011&   1.9296&   1.4447\\
 $5d^2\ ^1G_4$&$5d6p\ ^3F_4$&  0.1544&  0.1571&   0.1535&   0.1114\\
\end{tabular}
\end{ruledtabular}
\label{tab-dip2}
\end{table}

\begin{figure*}
\centerline{\includegraphics[scale=0.30]{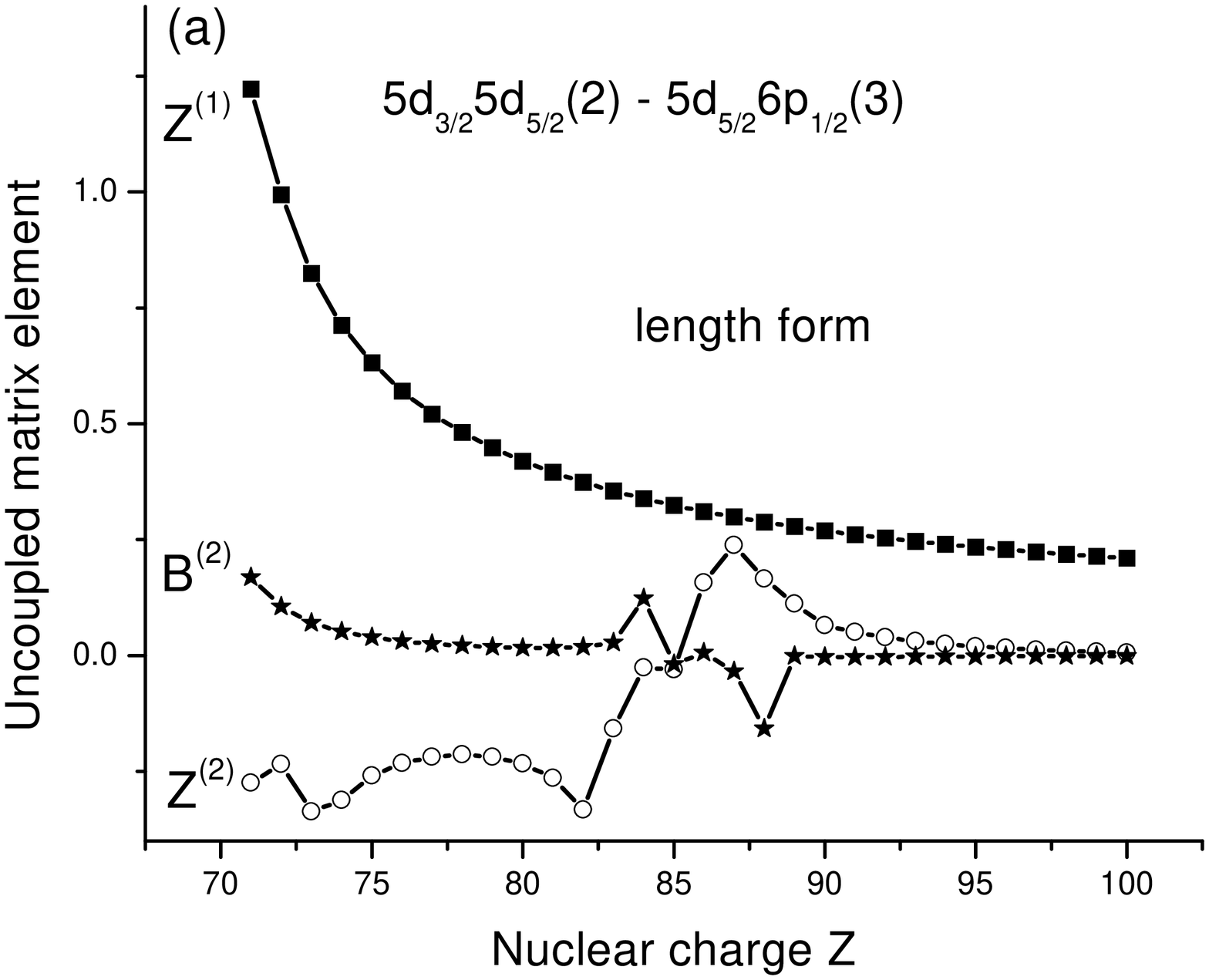}
\includegraphics[scale=0.30]{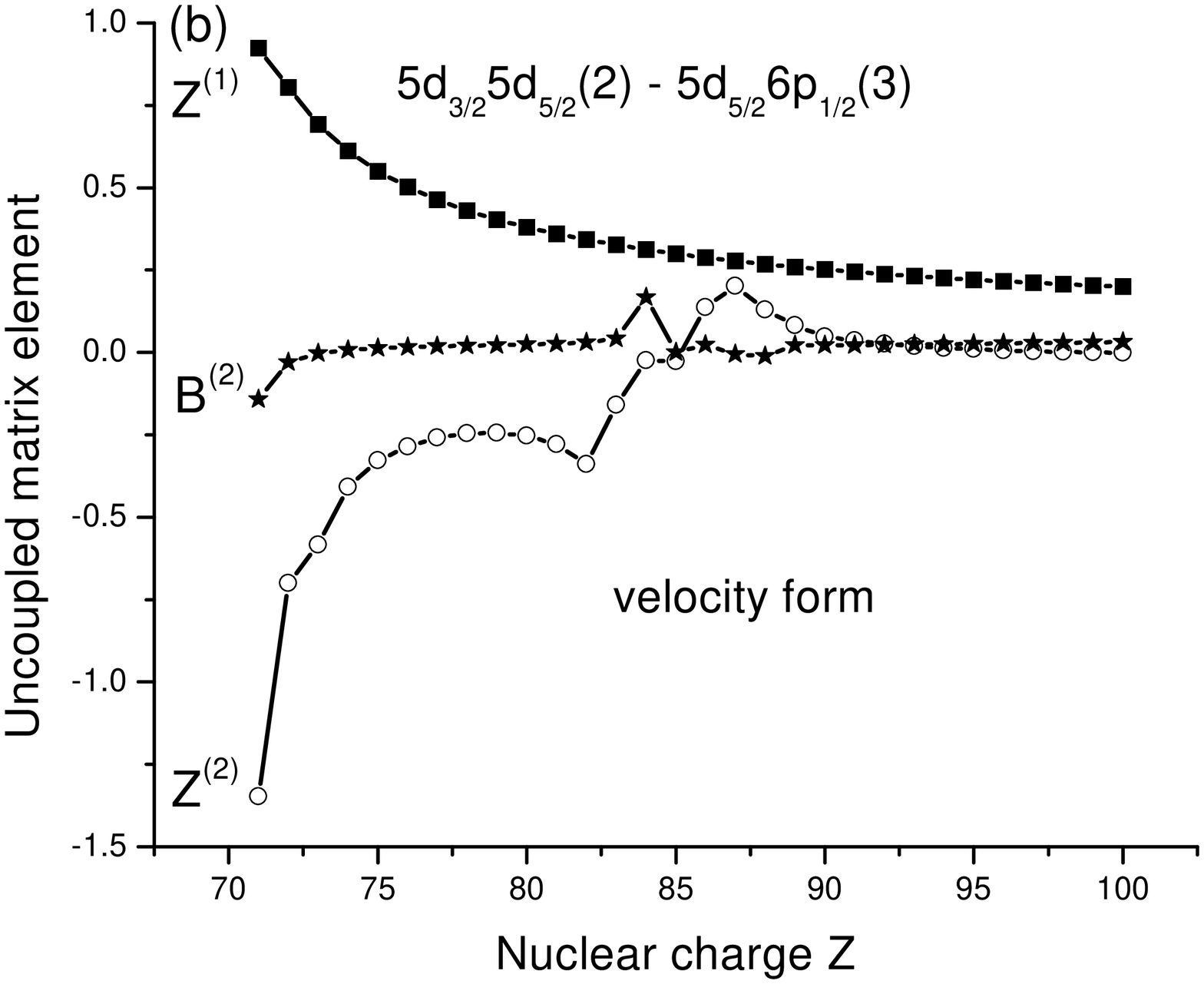}} \caption{Uncoupled
matrix elements for $5d_{3/2}5d_{5/2}(2)$-$5d_{5/2}6p_{1/2}(3)$
calculated in velocity and length forms}  \label{dip-lv}
\end{figure*}

\subsection{$Z$-dependence of matrix elements for
electric-dipole transitions}

We designate the first-order dipole matrix element by $Z^{(1)}$,
the Coulomb correction to the second-order matrix element by
$Z^{(2)}$, and the second-order Breit correction by $B^{(2)}$. The
evaluation of $Z^{(1)}$, $Z^{(2)}$, and $B^{(2)}$ for Yb-like ions
follows the pattern of the corresponding calculation for
beryllium-like ions in Refs.~\cite{be2t,be3t}. These matrix
elements are calculated in both length and velocity gauges. 
Differences between length and velocity forms, are illustrated
for the uncoupled $5d_{3/2}5d_{5/2}(2)$-$5d_{5/2}6p_{1/2}(3)$
matrix element in Fig.~\ref{dip-lv}. The
second-order Breit matrix element $B^{(2)}$ is multiplied by a factor of 50 
 in order to put it on the same scale as the second-order 
Coulomb matrix element $Z^{(2)}$. The sharp
 features have the same origin as those in the
the second-order energy matrix and are discussed in Appendix B.
Contributions of the second-order matrix elements $Z^{(2)}$ and
$B^{(2)}$ are much larger in velocity ($V$) form than in length ($L$) form
as seen  in panels $(a)$ and $(b)$ of Fig.~\ref{dip-lv}. 
As shown later, $L$--$V$ differences are compensated by
``derivative terms'' $P^{({\rm derv})}$.

\subsection{Example: dipole matrix elements in Re$^{+5}$ }
 We list uncoupled first-
and second-order dipole matrix elements $Z^{(1)}$, $Z^{(2)}$,
$B^{(2)}$, together with derivative terms $P^{({\rm derv})}$ for
Re$^{+5}$ in Table~\ref{tab-dip}.
For simplicity, we consider only 
the 9 dipole transitions between even-parity states
with $J$=0  and odd-parity states with $J$=1.
The derivative terms shown in Table~\ref{tab-dip} arise because
transition amplitudes depend on energy, and the transition energy
changes order-by-order in perturbation theory. Both $L$
and $V$ forms are given for the matrix elements. We 
see that the first-order matrix elements $Z^{(1)}_L$ and
$Z^{(1)}_V$ differ by 10-20\% and that the $L$--$V$ differences between
second-order matrix elements are much larger for some transitions.
The first-order matrix elements $Z^{(1)}_L$ and
$Z^{(1)}_V$ are non-zero for the three matrix elements
$5d_{3/2}5d_{3/2}(0)$--$5d_{3/2}6p_{1/2}(1)$,
$5d_{3/2}5d_{3/2}(0)$--$5d_{3/2}6p_{3/2}(1)$, and
$5d_{5/2}5d_{5/2}(0)$--$5d_{5/2}6p_{3/2}(1)$, but vanish
for the remaining six transitions.
Moreover, those six transitions, 
$Z^{(2)}_L$ and $Z^{(2)}_V$ have the same order of magnitude as 
the three non-zero first-order matrix elements.
This confirms the importance of including higher-order RMBPT 
corrections. It can also see from
Table~\ref{tab-dip} that $P^{({\rm derv)}}_{L}$  is almost
equal to $Z^{(1)}_{L}$  but  $P^{({\rm derv)}}_{V}$
is smaller than $Z^{(1)}_{V}$ by five to six orders of magnitude.

In Table \ref{tab-dip2}, line strengths for Re$^{+5}$ 
in length and velocity forms are
given for the $J$=0--$J'$=1 
transitions considered in Table~\ref{tab-dip} as well as for 
other $J$--$J'$ transitions. We see that $L$ and $V$ forms of the
coupled matrix elements in Table~\ref{tab-dip2} differ only in
the second or third digits. As mentioned in the introduction, 
these $L$--$V$ differences arise
because we start our RMBPT calculations using a non-local
DF potential. If we were to replace the DF potential
by a local potential, the differences would disappear completely.
Another source of $L$--$V$ differences is the use of an incomplete
model space.   The last two columns in
Table \ref{tab-dip2} show $L$ and $V$ values of line strengths
calculated without the second-order
contribution. As can be seen, including second-order corrections significantly 
decreases $L$--$V$ differences.

\section{Results and discussion}

We calculate energies of the 14 even-parity $5d^2$, $5d6s$, and
$6s^2$ states as well as  the 36 odd-parity $5d6p$, $5d5f$, and $6s6p$ states
for Yb-like ions with nuclear charges in the range $Z$ = 72
to 100.  Reduced matrix elements, line strengths, oscillator
strengths, and transition rates are also determined for all
electric-dipole transitions between even-parity and  
odd-parity  states for
each ion. Comparisons  with  experimental data and other theoretical
results are also given. Our results are presented in
three parts: transition energies, fine-structure energy
differences, and trends of line strengths, oscillator strengths,
and  transition rates.

\begin{table*}
\caption{ Energy levels (cm$^{-1}$) in ytterbium isoelectronic
sequence. Comparison of RMBPT  results with experimental data
presented in Refs.~\protect\cite{sugar,hof,kila,kilb,gayasov}. }
\begin{ruledtabular}
\begin{tabular}{lrrrrrrrrrrrr}
\multicolumn{1}{c}{Level} & \multicolumn{2}{c}{Re~VI} &
\multicolumn{2}{c}{Os~VII} & \multicolumn{2}{c}{Ir~VIII} &
\multicolumn{2}{c}{Pt~IX} & \multicolumn{2}{c}{Au~X} &
\multicolumn{2}{c}{ Hg~XI}\\
\multicolumn{1}{c}{} & \multicolumn{1}{c}{RMBPT} &
\multicolumn{1}{c}{\protect\cite{sugar}} &
\multicolumn{1}{c}{RMBPT} &
\multicolumn{1}{c}{\protect\cite{hof,kila}} &
\multicolumn{1}{c}{RMBPT} &
\multicolumn{1}{c}{\protect\cite{hof,kila}} &
\multicolumn{1}{c}{RMBPT} &
\multicolumn{1}{c}{\protect\cite{kilb}} &
\multicolumn{1}{c}{RMBPT} &
\multicolumn{1}{c}{\protect\cite{gayasov}} &
\multicolumn{1}{c}{RMBPT} &
\multicolumn{1}{c}{\protect\cite{gayasov}}\\
\hline
$5d^2\ ^3F_2 $&     0  &     0    &    0   &   0     &   0   &    0    &   0    &  0      &  0    &  0       &  0    &  0    \\
$5d^2\ ^3F_3 $&    8491&    8167  &   10604&  10308  &  12932&   12672 &   15490&  15250  &  18284&  18077   &  21325&  21175\\
$5d^2\ ^3P_0 $&   14495&   14379  &   16117&  15837  &  17670&   17254 &   19174&  18612  &  20638&  19949   &  22055&  21255\\
$5d^2\ ^3F_4 $&   15095&   14679  &   18409&  18049  &  21920&   21615 &   25633&  25359  &  29550&  29309   &  33677&  33481\\
$5d^2\ ^1D_2 $&   16866&   16577  &   19852&  19444  &  22905&   22436 &   26146&  25568  &  29568&  28925   &  33211&  32513\\
$5d^2\ ^3P_1 $&   19556&   19142  &   22624&  22098  &  25841&   25222 &   29238&  28515  &  32832&  32039   &  36643&  35802\\
$5d^2\ ^1G_4 $&   26807&   26657  &   31496&  31274  &  36522&   36253 &   41949&  41610  &  47818&  47451   &  54165&  53918\\
$5d^2\ ^3P_2 $&   28371&   27723  &   33932&  33127  &  39735&   38878 &   45962&  44991  &  52588&  51566   &  59691&  58835\\
$5d^2\ ^1S_0 $&   51056&   50492  &   58851&  57710  &  66656&   65022 &   74624&  72527  &  82848&  80385   &  91380&  88897\\
$5d6s\ ^3D_1 $&   92656&   92312  &  129438& 129244  & 169138&  169223 &  211579&         & 256622&          & 304159&       \\
$5d6s\ ^3D_2 $&   94554&   94262  &  131488& 131416  & 171413&  171485 &  214009&         & 259236&          & 306918&       \\
$5d6s\ ^1D_2 $&  109374&  108709  &  148595& 148023  & 191018&  190553 &  236412&         & 284633&          & 335596&       \\
$5d6s\ ^3D_3 $&  103444&  102712  &  142679& 142163  & 185041&  184748 &  230364&         & 278518&          & 329409&       \\
$6s^2\ ^1S_0 $&  198652&          &  272504& 282345  & 352159&  363633 &  437184&         & 527172&          & 621600&       \\
$5d6p\ ^3F_2 $&  163618&  162134  &  209906& 208659  & 259044&  258017 &  310888& 310064  & 365322& 364711   & 422250& 422506\\
$5d6p\ ^3D_1 $&  166538&  166077  &  213084& 212862  & 262439&  262461 &  314456& 314751  & 369016& 369624   & 426038& 426672\\
$5d6p\ ^3D_2 $&  176168&  174761  &  224886& 223453  & 276670&  275872 &  331369& 330850  & 388865& 388660   & 449056& 449335\\
$5d6p\ ^3F_3 $&  177212&  175502  &  226212& 224741  & 278237&  276994 &  333148& 332117  & 390829& 390038   & 451183& 450786\\
$5d6p\ ^1D_2 $&  182853&  181154  &  233874& 232453  & 288280&  287122 &  345942& 345026  & 406770& 406109   & 470688& 470408\\
$5d6p\ ^3D_3 $&  186672&  185022  &  238186& 236726  & 293068&  291792 &  351186& 350075  & 412435& 411533   & 476719& 476134\\
$5d6p\ ^3P_1 $&  187037&  185963  &  238656& 237682  & 293545&  292734 &  351634& 350986  & 412843& 412413   & 477109& 476749\\
$5d6p\ ^3P_0 $&  189199&  187780  &  241043& 239794  & 296204&  295142 &  354580& 353689  & 416091&          & 480682& 480075\\
$5d6p\ ^3F_4 $&  195446&  193260  &  249334& 247354  & 306792&  305005 &  367714& 366087  & 432023& 430576   & 499662& 498310\\
$5d6p\ ^3P_2 $&  196486&  194539  &  250142& 248453  & 307362&  305894 &  368050& 366803  & 432129& 431126   & 499533& 498771\\
$5d6p\ ^1F_3 $&  197742&  195691  &  251254& 249401  & 308315&  306697 &  368814& 367435  & 432644& 431579   & 499643& 499109\\
$5d6p\ ^1P_1 $&  200506&  200437  &  255391& 255246  & 313586&  313520 &  375052& 375145  & 439684& 440120   & 507190& 508193\\
$6s6p\ ^3P_0 $&  261637&          &  343490&         & 408884&         &  460493& 456143  & 510225&          & 558566&       \\
$6s6p\ ^3P_1 $&  265550&          &  341677& 341844  & 394150&  391888 &  443224& 438038  & 490445&          & 536100&       \\
$6s6p\ ^3P_2 $&  281644&  279156  &  333322& 332501  & 381777&  379116 &  428519& 424403  & 473776&          & 517742&       \\
$5d5f\ ^1G_4 $&  282611&  280575  &  333410& 331454  & 382478&  380334 &  430103& 425349  & 476544&          & 522022&       \\
$5d5f\ ^3F_2 $&  282986&  281906  &  341072& 340330  & 391578&  388229 &  440422& 435302  & 487930&          & 534343&       \\
$5d5f\ ^3F_3 $&  283375&  283559  &  334390& 334240  & 383305&  380724 &  430253& 422093  & 475351&          & 518717&       \\
$5d5f\ ^3H_4 $&  284803&  282853  &  336142& 334302  & 385817&  383508 &  434072& 430412  & 481126&          & 527163&       \\
$5d5f\ ^3G_3 $&  286237&  286426  &  338249& 337436  & 388358&  385282 &  436878& 425627  & 484128&          & 530397&       \\
$5d5f\ ^3H_5 $&  287360&  285346  &  339706& 337535  & 390378&  387545 &  439680& 436230  & 487868&          & 535150&       \\
$5d5f\ ^3D_1 $&  289100&  290208  &  349008&         & 408218&  406542 &  459678& 456053  & 509418&          & 557828&       \\
$5d5f\ ^1D_2 $&  289565&  288853  &  348392& 347386  & 399983&  397557 &  449925& 442048  & 498451&          & 545750&       \\
$5d5f\ ^3F_4 $&  292767&  290824  &  345648& 343897  & 396931&  393669 &  446783& 439406  & 495294&          & 542569&       \\
$5d5f\ ^3G_4 $&  294002&  293741  &  347287& 346144  & 398865&  396001 &  449051& 445365  & 498219&          & 546694&       \\
$5d5f\ ^3D_2 $&  295265&  294688  &  354838& 354012  & 408411&  405867 &  460122& 456798  & 510546&          & 559987&       \\
$6s6p\ ^1P_1 $&  295457&          &  357819& 356849  & 420585&  421774 &  477574&         & 532630&          & 586680&       \\
$5d5f\ ^3D_3 $&  295676&  295836  &  349683& 349155  & 401804&  397003 &  452259& 436157  & 501264&          & 549026&       \\
$5d5f\ ^3G_5 $&  295717&  295928  &  349829& 349324  & 402320&  399128 &  453392& 443691  & 503255&          & 552106&       \\
$5d5f\ ^3H_6 $&  296143&  292600  &  349940&         & 402394&         &  453792&         & 504385&          & 554394&       \\
$5d5f\ ^1F_3 $&  300071&  300794  &  356198& 356322  & 410808&  408418 &  464085& 461735  & 516269&          & 567655&       \\
$5d5f\ ^3P_2 $&  301078&  301178  &  371823&         & 465428&         &  565480&         & 671378&          & 782904&       \\
$5d5f\ ^3P_1 $&  301570&  302600  &  362544& 363927  & 440737&         &  534533&         & 634090&          & 738793&       \\
$5d5f\ ^3P_0 $&  302331&  303600  &  358117& 356587  & 434121&         &  527597&         & 626622&          & 730708&       \\
$5d5f\ ^1H_5 $&  303559&  307440  &  361850& 364949  & 418559&  420291 &  473757& 470214  & 527612&          & 580323&       \\
$5d5f\ ^1P_2 $&  317422&  317600  &  396700&         & 489065&         &  589085&         & 695429&          & 807602&       \\
\end{tabular}
\end{ruledtabular} \label{tab-com}
\end{table*}

\subsection{Transition energies}

In Table \ref{tab-com}, transition energies are compared with
recent measurements \cite{sugar,hof,kila,kilb,gayasov,churilov,optk}. 
We obtain good agreement with the experiment for 
low-lying   $5d^2$, $5d6s$, and $5d6p$  levels.
We also  we obtain reasonable agreement (500 -1000 cm$^{-1}$) 
for highly-excited  $5d5f$ levels in Re~VI, Os~VII, and Ir~VIII 
ions. However,  we disagree substantially (2000 -10000 cm$^{-1}$)
for $5d5f$ levels in Pt~IX, Au~X, and Hg~XI ions.
As mentioned in Refs.~\cite{kilb,gayasov,churilov}, 
mixing between levels of the $5d5f$ configuration and 
the core excited configurations 
with a $5p$ hole could be very important for Pt~IX, Au~X, and Hg~XI ions.  
Indeed, substantial mixing of the $5p^55d^3$ and
$5p^65d5f$ configurations was found in Ref.~\cite{churilov}.
From this, we conclude that the present values for $5d5f$ configurations
in Pt~IX, Au~X, and Hg~XI are less reliable ($\sim$ 1\%) than data 
for Re~VI, Os~VII, and Ir~VIII.

\begin{figure*}[tbp]
\centerline{\includegraphics[scale=0.30]{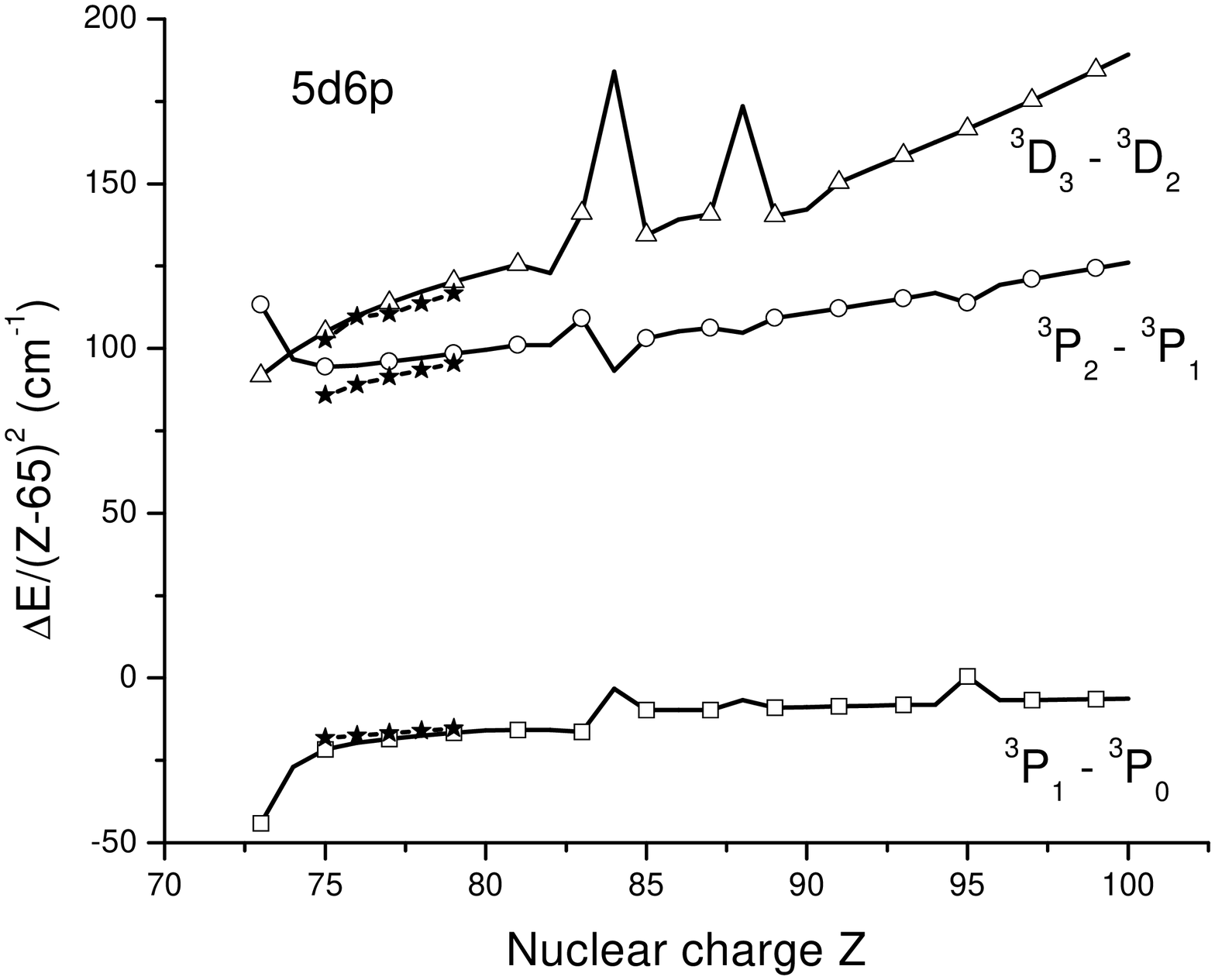}
            \includegraphics[scale=0.30]{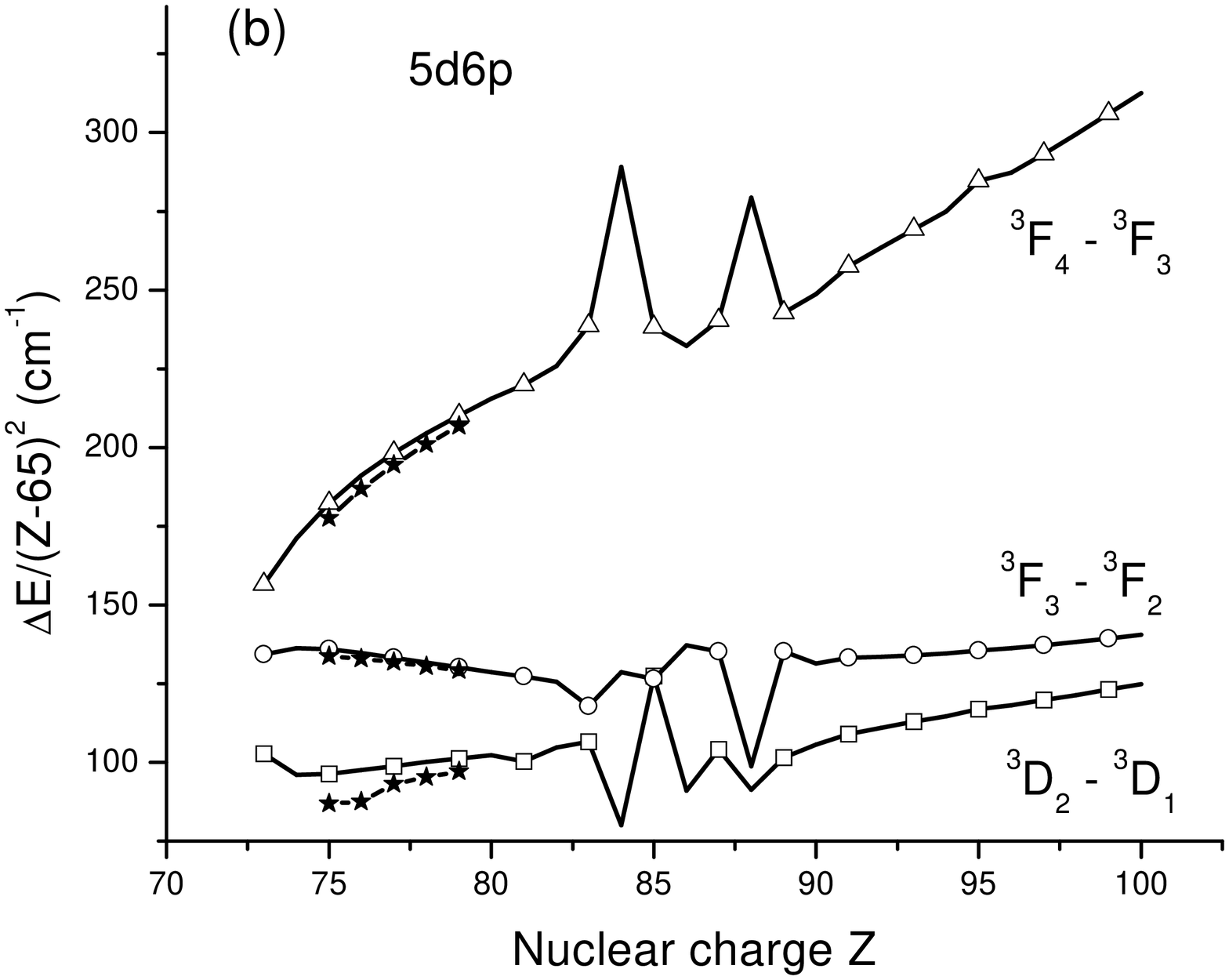}}
\caption{Fins-structure intervals $\Delta E/(Z-65)^2$ in cm$^{-1}$ of
$5d6p\ ^3P$, $5d6p\ ^3D$, and $5d6p\ ^3F$ terms as function of
$Z$. Experimental data represented by the $\ast$ symbol are from
\protect \cite{sugar,hof,kila,kilb,gayasov,churilov,optk}} \label{spl-pd}
\end{figure*}

\subsection{Fine structure of the  $5d6p$ triplets}

The fine-structure intervals for the $^3P$, $^3D$, and $^3F$ terms of
$5d6p$ configuration divided by $(Z-65)^2$ are shown in Fig.~\ref{spl-pd}.
The fine structures of these levels do not follow the Land\'{e} rules
even for small $Z$;  the $^3P$  levels are partially inverted,
while the $^3D$ and $^3F$ levels show regular ordering of the
fine-structure splittings for both low and high $Z$. The unusual
splittings are caused by changes from $LS$ to $jj$
coupling and by mixing from other triplet and singlet states.
Comparisons are made with experimental data from
\cite{sugar,hof,kila,kilb,gayasov,churilov,optk} 
 in Fig.~\ref{spl-pd}. 
Excellent agreement (0.3\% -- 3\%) is found for  
the four intervals $^3P_1$--$^3P_0$,
$^3D_3$--$^3D_2$, $^3F_4$--$^3F_3$, and $^3F_3$--$^3F_2$.
The agreement decreases for the $^3P_2$--$^3P_1$ 
and $^3D_2$--$^3D_1$ intervals, from 10\% for Re$^{+5}$ to 2\% for Hg$^{+10}$.
Experimental energies for other ions would be very helpful 
in confirming the $Z$ dependence shown in Fig.~\ref{spl-pd}.

\begin{figure*}[tbp]
\centerline{\includegraphics[scale=0.30]{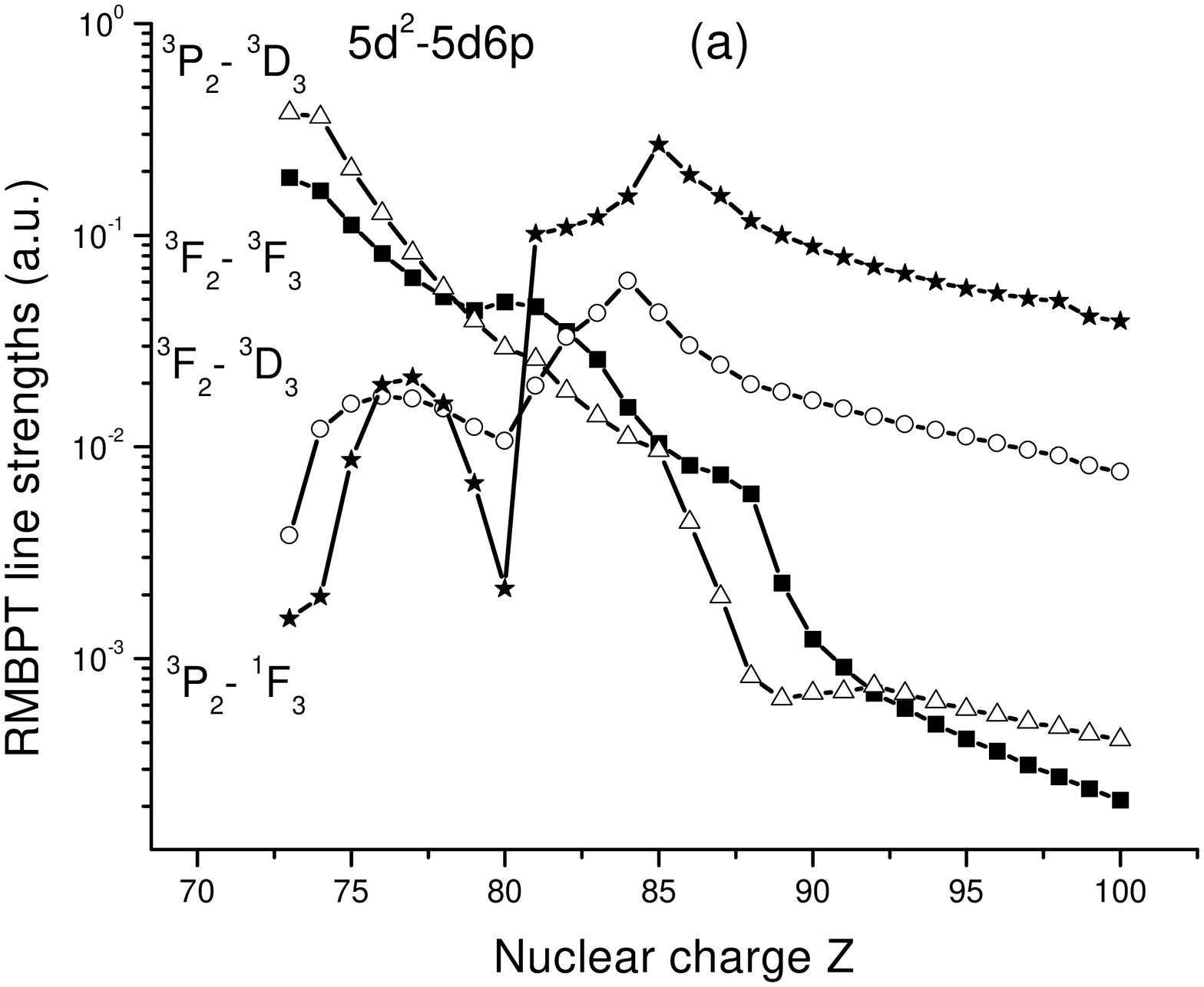}
\includegraphics[scale=0.30]{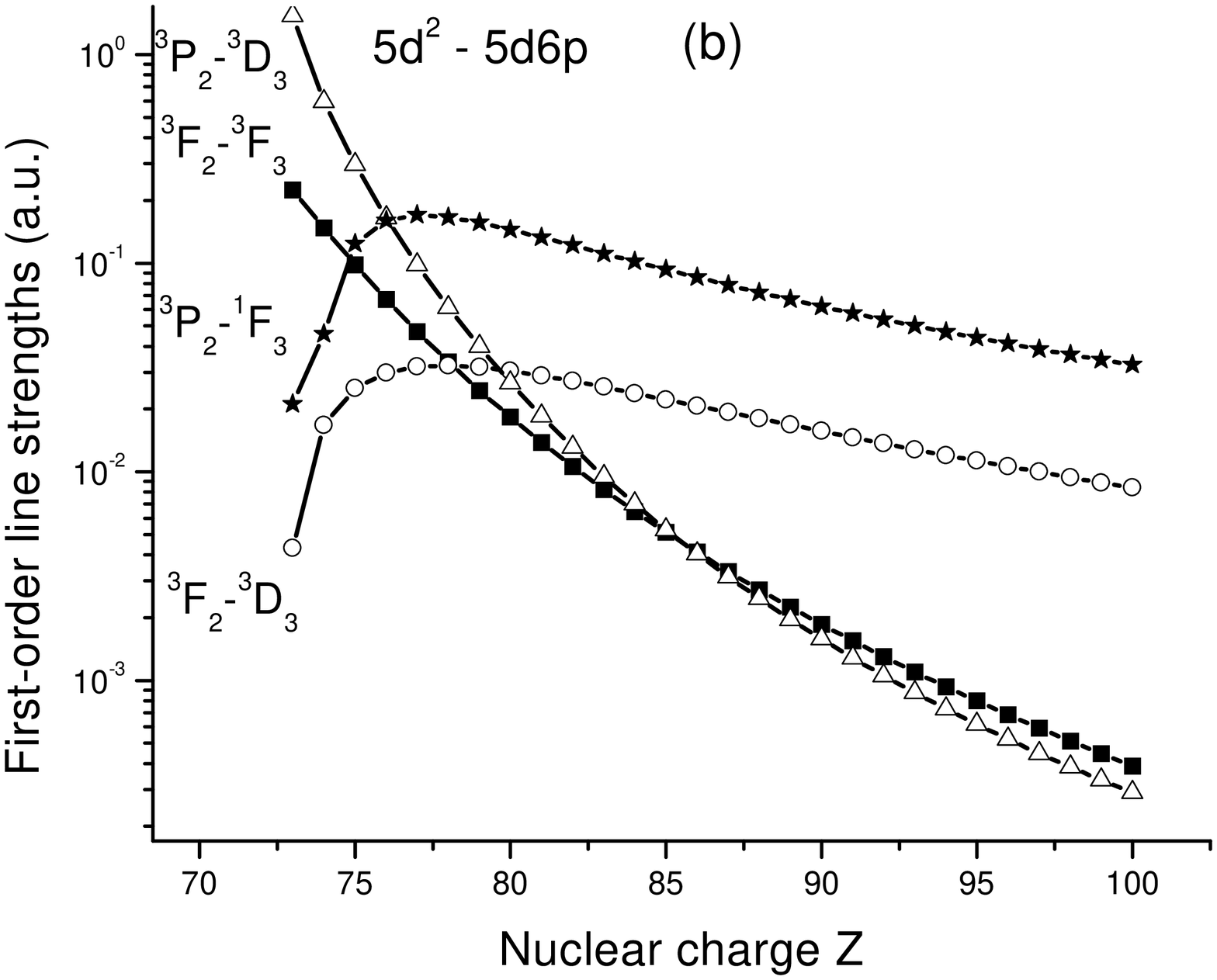}} \caption{ Line strengths
for $5d^2-5d6p$ transitions in Yb-like ions: (a) - RMBPT, (b) - First-order
approximation.} \label{fig-line}
\end{figure*}

\begin{figure*}[tbp]
\centerline{\includegraphics[scale=0.30]{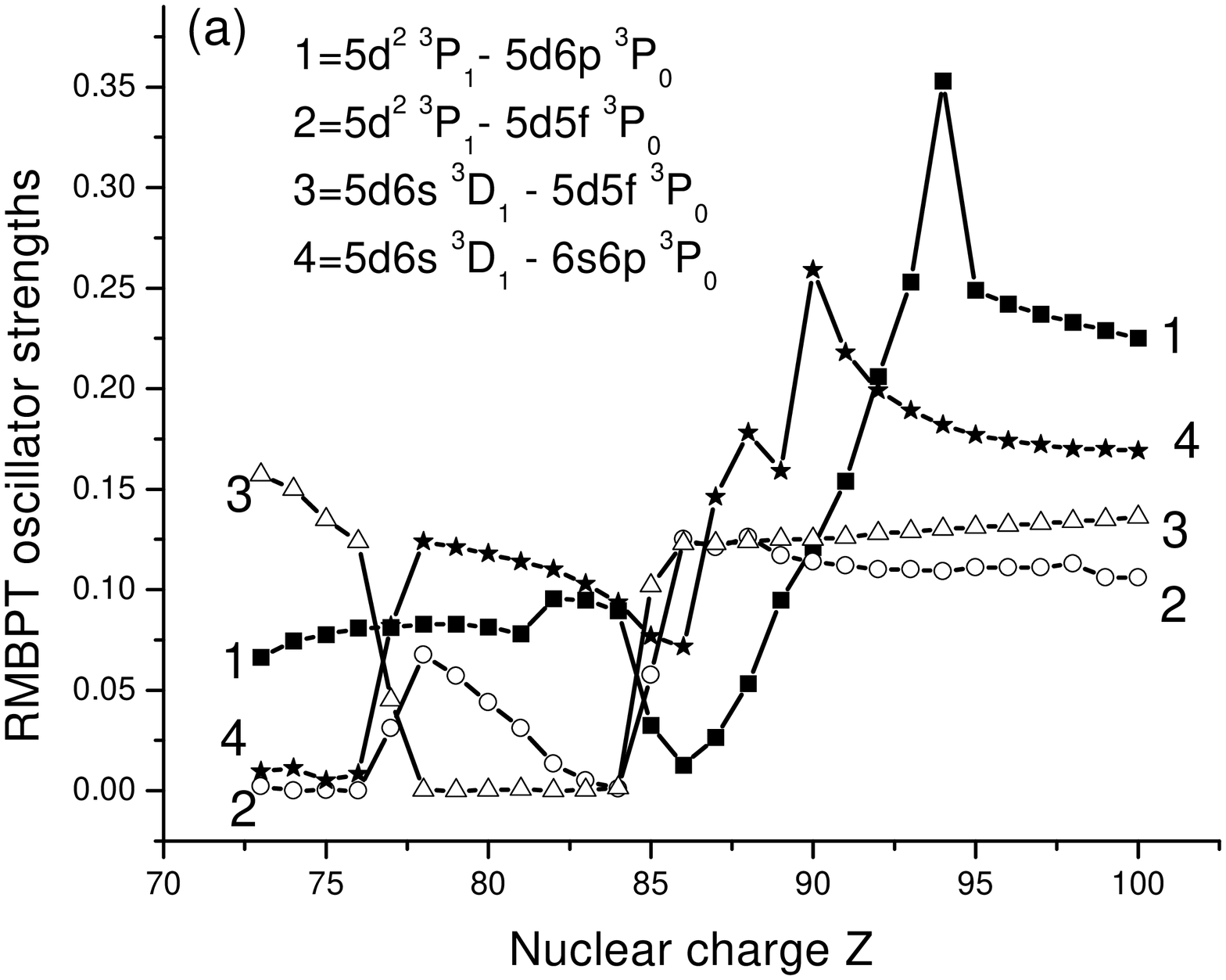}
            \includegraphics[scale=0.30]{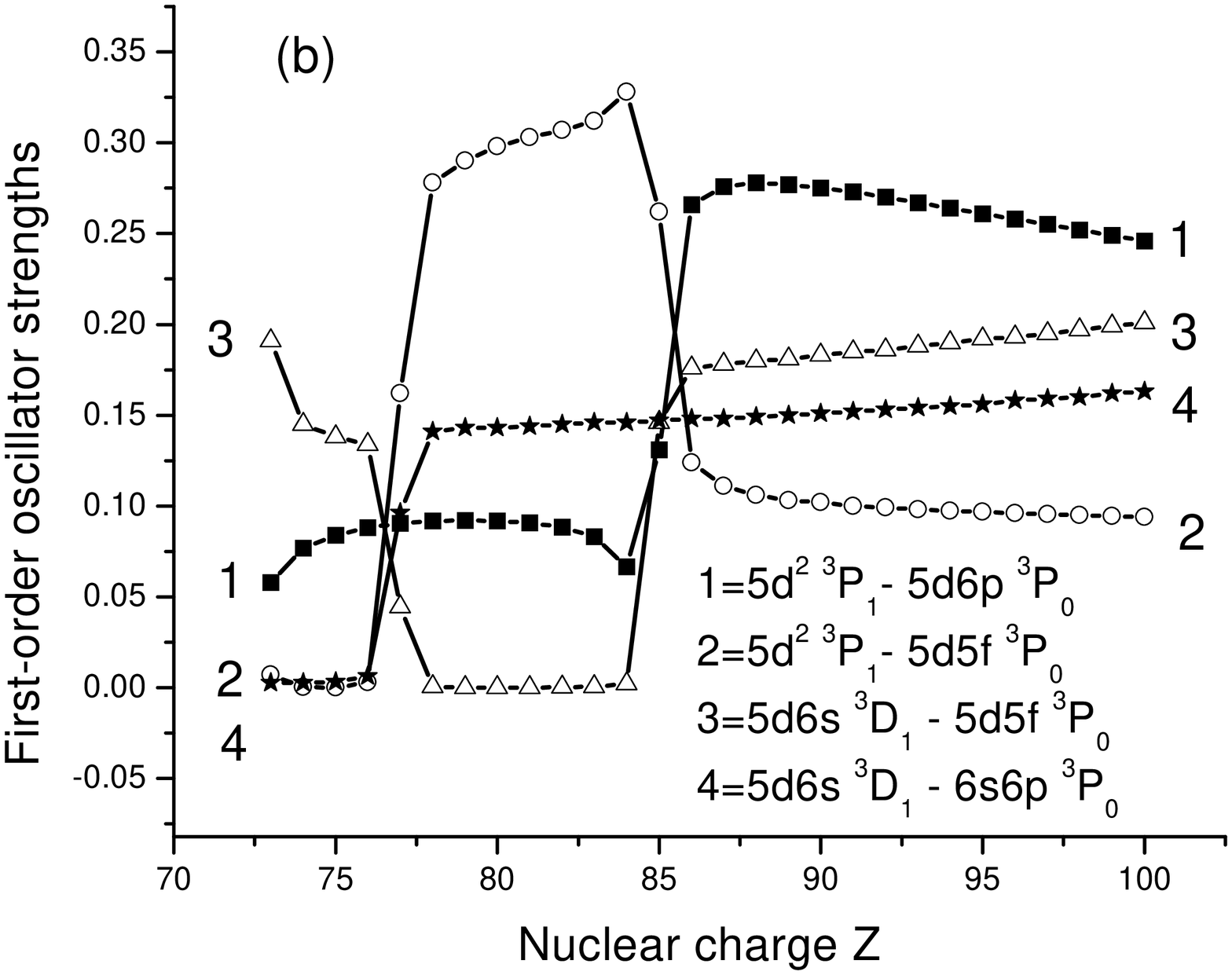}} \caption{
Oscillator strengths for transitions from even-parity complex
with $J$=1 to odd-parity complex with $J$=0  in Yb-like ions: (
a) - RMBPT, (b) - First-order
approximation.}
\label{fig-osc}
\end{figure*}

\subsection{Line strengths and oscillator strengths
 in Yb-like ions}

Trends of the $Z$ dependence of line strengths and oscillator strengths,
are shown in Figs.~\ref{fig-line} and \ref{fig-osc}.
In Fig.~\ref{fig-line}, we illustrate the $Z$ dependence
of  line strengths
for the four $5d^2$--$5d6p$ transitions calculated (a) by RMBPT 
and (b) in first-order  to illustrate once again the importance of including
second-order matrix elements.
Comparing the curves in Fig.~\ref{fig-line}a and
Fig.~\ref{fig-line}b,
we can see that the  $Z$ dependence is the same for both cases
except for the small region  $Z$ = 80--88. 
The singularities in this region are already present for uncoupled
dipole matrix elements.
The  second-order contribution for the
$5d_{3/2}5d_{5/2}(2)$--$5d_{5/2}6p_{1/2}(3)$ uncoupled
matrix element is shown in Fig.~\ref{dip-lv}. Similar singularities
appeared for other uncoupled matrix elements between even-parity $J$=2 
states and odd-parity $J$=3 states. Consequently, 
those singularities also appear in the coupled matrix elements and 
the line strengths shown in Fig.~\ref{fig-line}.

    The $Z$ dependence is more complicated for transitions
between even-parity states with $J$=1 and odd-parity
states with $J$=0, as seen in Fig.~\ref{fig-osc}, where we give
oscillator strengths for four transitions calculated (a) by RMBPT and (b)
in first-order. The oscillator strength curves are 
not smooth functions of $Z$, even in first-order.
The singularities shown in the four curves of Fig.~\ref{fig-osc}b 
arise from the strong
mixing between three states of the odd-parity complex with $J$=0.
The mixing of $5d_{5/2}5d_{5/2}$, $5d_{3/2}6p_{3/2}$, and $6s_{1/2}6p_{1/2}$
states was discussed earlier in Fig.~\ref{en-odd}a
where it was shown that avoided crossings occur near $Z$=76 and 84.

\section{Conclusion}
In summary,  a systematic second-order RMBPT study of the energies
of the  $5d^2$, $5d6s$, $6s^2$, $5d6p$, $5d5f$, and $6s6p$ states
of Yb-like ions has been presented. These calculations 
agree with existing experimental energy data for intermediate
$Z$=75 to 80 at the level of 500-1000~cm$^{-1}$ for low-lying
$5d^2$, $5d6s$, and $5d6p$ levels. They provide a smooth
theoretical reference database for the identification of lines.

Also presented is a systematic second-order relativistic RMBPT
study of reduced matrix elements, line strengths, oscillator
strengths, and transition rates for allowed and forbidden electric-dipole
transitions in Yb-like
ions  with nuclear charges ranging from $Z$ = 72 to 100. The
dipole matrix elements include retardation and correlation corrections from
Coulomb and Breit interactions.  Both length and velocity forms of
the matrix elements are evaluated, and $\sim 5$~\% differences, caused
by the non-locality of the starting DF potential, are found
between the two forms. 

 We believe that our results will be useful in analyzing existing
experimental data and planning new experiments. There remains a
paucity of experimental data for many of the higher ionized
members of this sequence, both for term energies and for
transition probabilities and lifetimes. Additionally, matrix
elements from the present calculations will provide basic
theoretical input for calculations of reduced matrix elements,
oscillator strengths, and transition rates in three-electron
Lu-like ions.

\begin{acknowledgments}
The work of W.R.J. and M.S.S. was supported in part by National
Science Foundation Grant No.\ PHY-01-39928. U.I.S. acknowledges
partial support by Grant No.\ B503968 from Lawrence Livermore
National Laboratory. The work of J.R.A was performed under the
auspices of the U. S. Department of Energy by the University of
California, Lawrence Livermore National Laboratory under contract
No.\ W-7405-Eng-48. We also acknowledge helpful discussions with
Professor A.N. Ryabtsev.
\end{acknowledgments}

\appendix

\section{The second-order double-excitation  contribution}

Here, we consider the second-order double-excitation contribution
to explain singularities in the $Z$ dependence of energy levels given in
Figs.~\ref{en-eve} and \ref{en-odd}. A typical
contribution from one of the double-excitation diagram for the
second-order interaction energy has the form \cite{be2e}
\begin{equation}\label{eq1}
A_{l}(vw;v'w')
\propto \sum_{mn}\sum_{kk^{\prime }}\frac{X_{k}(vwmn)X_{k^{\prime }}(mnv'w')}
{\epsilon _{v'}+\epsilon _{w'}-\epsilon
_{m}-\epsilon _{n}}\,,
\end{equation}
where $X_{k'}(vwmn)$ are products of angular coupling coefficients and
Slater integrals \cite{be2e},
$\epsilon_m$ is the single-particle energy for state $m$,
 and $v$ designates a
single-particle valence state ($n_vl_vj_v$).
For the case of a [Xe]$4f^{14}$ core,
sums  over $m$ and $n$ include $5d_{3/2}$, $5d_{5/2}$, $5f_{5/2}$,
$5f_{7/2}$, $5g_{7/2}$, $5g_{9/2}$, and all states with principal
quantum number $n>$5. We must, however, 
exclude terms with pairs ($mn$) that are included in model space.
Therefore, 
we remove the pairs ($5d5d$), ($5d6s$), and ($6s6s$) since they are
in the even-parity model space and pairs ($5d6p$), ($5d5f$), and
($6s6p$) in the odd-parity model space.

As an example, let us consider the singularity near $Z$=85 in the curve 
describing the energy of the $6s^2$ level in Fig.~\ref{en-eve}b. 
The denominator in Eq.~(\ref{eq1}) is 
$D= 2\epsilon_{6s}-\epsilon_m-\epsilon_n$ for 
the matrix element  $A_l(6s_{1/2}6s_{1/2};6s_{1/2}6s_{1/2})$.
For an ion with
$Z$=85, $D$ becomes very small when
($mn$)=($5d_{3/2}6d_{3/2}$)\, since
${\epsilon_{6s_{1/2}}}$=-9.488404 a.u.,
${\epsilon_{5d_{3/2}}}$=-12.069226 a.u., and
${\epsilon_{6d_{3/2}}}$=-6.932272 a.u., giving
$D$=0.02469.  This one term dominates the  matrix element
$A_l(6s_{1/2}6s_{1/2};6s_{1/2}6s_{1/2})$  and increases its 
size by a factor of six in comparison with values 
for neighboring $Z$, explaining the singulatity at $Z$=85 in Fig.~\ref{en-eve}b.
The explanation of the sharp features in the curves describing
energies of the $6s_{1/2}6p_{1/2}$ and $5d_{5/2}5f_{5/2}$ levels in
Fig.~\ref{en-odd} is similar.

\section{The second-order dipole matrix element}
A typical contribution from one of the second-order correlation
corrections to the dipole matrix element ($vw(J)-v'w'(J')$ has the
form \cite{be2t}
\[\label{b1}
Z^{(\text{corr})}[vw(J)-v^{\prime }w^{\prime }(J^{\prime })]\propto \sum_{i}%
\frac{Z_{iv}X_{k}(v^{\prime }w^{\prime }wi)}{\epsilon
_{i}+\epsilon _{w}-\epsilon _{v^{\prime }}-\epsilon _{w^{\prime
}}}\, ,
\]
where  $Z_{iv}$ is a single-electron dipole matrix element. 
In the sum over $i$, only terms with vanishing
denominators are excluded.
 For the 
$5d_{5/2}5d_{3/2}(2)-5d_{5/2}6p_{1/2}(3)$ transition, we obtain
\begin{eqnarray*}\label{b2}
\lefteqn{Z^{(\text{corr})}[5d_{5/2}5d_{3/2}(2)-\
5d_{5/2}6p_{1/2}(1)]}\hspace{4em}
\nonumber \\
&&\propto \sum_{i}\frac{Z(i,5d_{5/2})X_{k}(5d_{5/2}6p_{1/2}\ 5d_{3/2}i)}{%
\epsilon _{i}+\epsilon (5d_{3/2})-\epsilon (5d_{5/2})-\epsilon
(6p_{1/2})}\, .
\end{eqnarray*}
For the case  $i$=$5f_{7/2}$ and  $Z$=84, the
denominator becomes
\[\label{b3}
\epsilon _{i}+\epsilon (5d_{3/2})-\epsilon (5d_{5/2})-\epsilon
(6p_{1/2}) =0.003.
\]
 This one term dominates the entire matrix element.
For the 
$5d_{5/2}5d_{5/2}(0)-5d_{3/2}6p_{1/2}(1)$ transition, the
denominator becomes very small at $Z$=88 when $i$=$5f_{7/2}$.
\[\label{b4}
\epsilon _{i}+\epsilon (5d_{5/2})-\epsilon (5d_{3/2})-\epsilon
(6p_{1/2}) = -0.056.
\]
Again, this single term dominates the matrix element.

A typical contributions from the second-order RPA
correction for dipole matrix element ($vw(J)-v'w'(J')$ has the
form 
\[
Z^{{(\rm RPA)}}_{1}[vw(J)-v^{\prime }w^{\prime }(J^{\prime })]\propto \sum_{i}%
\frac{Z_{nb}X_{k}(wnv'b)}{\epsilon _{n}+\epsilon _{w}-\epsilon
_{v'}-\epsilon _{b}}\, ,
\]
where the index $b$ designates core states and $n$ designates
an excited state. For the special case of the
$5d_{5/2}5d_{3/2}[2]-6p_{1/2}5d_{5/2}[1]$  transition, we obtain
\begin{eqnarray*}\label{b7}
\lefteqn{Z^{{(\rm RPA)}}_{1}[5d_{3/2}5d_{5/2}(2)-\ 6p_{1/2}5d_{5/2}(3)]}
\hspace{2em} \\
&& \propto \sum_{n}\sum_{b}\frac{Z(b,n)X_{k}(5d_{3/2}b\
6p_{1/2}n)}{\epsilon _{n}+\epsilon (5d_{3/2})-\epsilon
(6p_{1/2})-\epsilon _{b}}\, .
\end{eqnarray*}
In the case of $b$=$5p_{3/2}$ and $n$=$5d_{3/2}$  for nuclear
charge $Z$=88, the denominator becomes
\[\label{b8}
\epsilon (5d_{3/2})+\epsilon
(5d_{3/2})-\epsilon (6p_{1/2})-\epsilon (5p_{3/2})  =- 0.278.
\]
As before, the small value of the denominator leads to an anomalous increase in
the size of the RPA matrix element.

\end{document}